\documentclass[sigconf]{acmart}

\PassOptionsToPackage{hyphens}{url}
\usepackage{hyperref}

\usepackage{booktabs}
\usepackage{xspace}
\usepackage{amsmath}
\usepackage{graphicx}
\usepackage{xcolor,amsmath}
\usepackage{enumitem}
\usepackage{pifont}
\usepackage{adjustbox}
\usepackage{fdsymbol}
\usepackage{subcaption}

\usepackage[linewidth=1pt]{mdframed}
\usepackage{framed}
\usepackage[strict]{changepage}
\definecolor{formalshade}{rgb}{0.95,0.95,1}
\definecolor{darkblue}{rgb}{0.0, 0.0, 0.55}
\newenvironment{formal}{
  
  \MakeFramed{\advance\hsize-\width\FrameRestore}
  \noindent\hspace{-4.55pt}
  \begin{adjustwidth}{}{7pt}

}
{
  \end{adjustwidth}\endMakeFramed
}

\renewcommand\footnotetextcopyrightpermission[1]{}

\newcommand{\tild}{$\sim$}

\newcommand{\Tdot}{$\medblackcircle$}
\newcommand{\TDot}{$\medcircle$}

\newcommand{\Par}[1]{\noindent\textbf{#1.}}
\usepackage[linesnumbered,ruled]{algorithm2e}

\SetKwComment{Comment}{\color{green!50!black}// }{}
\newcommand{\var}{\texttt}
\SetKwProg{Function}{function}{}{}

\begin{abstract}
Unmanned Aerial Vehicles autonomously perform tasks with the use of state-of-the-art control algorithms. These control algorithms rely on the freshness and correctness of sensor readings. Incorrect control actions lead to catastrophic destabilization of the process.

In this work, we propose a multi-part \emph{Sensor Deprivation Attacks} (SDAs), aiming to stealthily impact process control via sensor reconfiguration. In the first part, the attacker will inject messages on local buses that connect to the sensor. The injected message reconfigures the sensors, e.g.,~to suspend the sensing.
In the second part, those manipulation primitives are selectively used to cause adversarial sensor values at the controller, transparently to the data consumer.
In the third part, the manipulated sensor values lead to unwanted control actions (e.g. a drone crash).
We experimentally investigate all three parts of our proposed attack. Our findings show that i)~reconfiguring sensors can have surprising effects on reported sensor values, and ii)~the attacker can stall the overall Kalman Filter state estimation, leading to a complete stop of control computations.  As a result, the UAV becomes destabilized, leading to a crash or significant deviation from its planned trajectory (over 30 meters). We also propose an attack synthesis methodology that optimizes the timing of these SDA manipulations, maximizing their impact. Notably, our results demonstrate that these SDAs evade detection by state-of-the-art UAV anomaly detectors.

Our work shows that attacks on sensors are not limited to continuously inducing random measurements, and demonstrate that sensor reconfiguration can completely stall the drone controller. In our experiments, state-of-the-art UAV controller software and countermeasures are unable to handle such manipulations. Hence, we also discuss new corresponding countermeasures.
\end{abstract}

\begin{document}
\title{Sensor Deprivation Attacks for Stealthy UAV Manipulation}

\author{Alessandro Erba}
\orcid{0000-0003-2631-8829}
\email{alessandro.erba@kit.edu}
\affiliation{
  \institution{KASTEL Security Research Labs,\\Karlsruhe Institute of Technology}
  \streetaddress{Am Fasanengarten 5}
  \city{Karlsruhe}
  \country{Germany}
  \postcode{76131}
}
\authornote{Part of this work was done while Alessandro Erba was with CISPA Helmholtz Center for Information Security and Saarbrücken Graduate School of Computer Science, Saarland University.}

\author{John H. Castellanos}
\orcid{0000-0003-3368-400X}
\email{john.castellanos@hitachienergy.com}
\affiliation{
  \institution{Hitachi Energy Research}
  \city{Mannheim}
  \country{Germany}
}
\authornote{Part of this work was done while John H. Castellanos was with CISPA Helmholtz Center for Information Security.}

\author{Sahil Sihag}
\orcid{0000-0001-8424-2602}
\email{sahil.sihag@cispa.de}
\affiliation{
  \institution{CISPA Helmholtz Center for Information Security}
  \streetaddress{Stuhlsatzenhaus 5}
  \city{Saarbr\"ucken}
  \country{Germany}
  \postcode{66123}
}

\author{Saman Zonouz}
\orcid{}
\email{saman.zonouz@gatech.edu}
\affiliation{
  \institution{Georgia Institute of Technology}
  \city{Atlanta}
  \country{USA}
  \postcode{30332}
}

\author{Nils Ole Tippenhauer}
\orcid{0000-0001-8424-2602}
\email{tippenhauer@cispa.de}
\affiliation{
  \institution{CISPA Helmholtz Center for Information Security}
  \streetaddress{Stuhlsatzenhaus 5}
  \city{Saarbr\"ucken}
  \country{Germany}
  \postcode{66123}
}

\newcommand{\attack}{Sensor Deprivation Attack\xspace}
\newcommand{\attacks}{Sensor Deprivation Attacks\xspace}
\newcommand{\attackShort}{SDA\xspace}
\newcommand{\attacksShort}{SDAs\xspace}

\author{}

\setcopyright{none}
\settopmatter{printacmref=false}
\pagestyle{plain}
\maketitle

\section{Introduction}

In modern society, Robotic Vehicles (RVs) such as Unmanned Aerial Vehicles (UAVs) are deployed to accomplish a number of tasks. UAVs have been used to deliver medicines~\cite{snouffer2022six} and tested for life-saving deliveries~\cite{forbes21organs}. UAVs are used for consumers' good delivery~\cite{primedrone,walmartdrone} and are deployed in military operations~\cite{conflict,conflict2}, often relying on open-source flight controllers~\cite{corvo}. Given their importance, security concerns were raised. Those concerns were confirmed by several recent papers that showed critical security issues in flight stacks~\cite{son15rocking,Sathaye2022GPS,jeong23unrocking,jang23paralyzing,schiller2023drone}.

In UAVs, sensors are generally continuously queried by the Micro Controller Unit (MCU) for the latest sensor readings. Sensors are configured to update their readings with fixed sampling rates (higher than the MCU query rates).
As sensors are in dedicated chips, they transmit the data to the MCU via the local bus system or analog signals.

So far, it is unclear \emph{what would happen to the drone control if the attacker can reduce the sensor update rate, or prevent any sensor data from being delivered? How can an air-gapped remote attacker affect the MCU-sensor connection to reduce the sensor update rate? By such limited but realistic malicious capabilities, can the attacker take deterministic control over the drone's flight operation as opposed to trivially causing a crash?}

Prior work in the CPS field proposed the use of GPS spoofing for remote UAV manipulation~\cite{Sathaye2022GPS}. By precisely and continuously injecting the spoofed GPS signal an attacker can take the victim's UAV. Alternatively, it was proposed to use intentional electromagnetic interference (IEMI) to remotely inject noise in the MCU-sensor bus and continuously perturb the sensor readings received at the UAV controller~\cite{jang23paralyzing}. By doing so the attacker can crash the drone. Continuous spoofing of victim vehicle sensor readings, moreover will trigger anomaly detection as the sensor readings will suddenly be noisy and inconsistent~\cite{erba22genericconcealment}.
\begin{figure}
    \centering
    \includegraphics[width=\columnwidth]{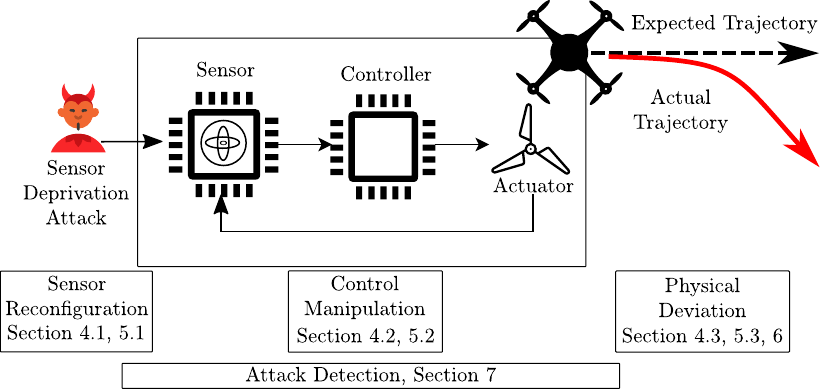}
    \caption{In a sensor deprivation attack, an adversary remotely reconfigures a flying drone's onboard sensor (e.g., via IEMI). The drone starts operating with the reconfigured sensor and makes wrong control decisions, which leads to deviation and crash. With a series of reconfigurations of the sensor, the attacker can even remotely control the drone.}

    \label{fig:enter-label}
\end{figure}

In this work, we introduce \attacks (\attacksShort) and investigate their impact on modern UAVs (see Figure~\ref{fig:enter-label}). \attacksShort are a new class of attacks that manipulate the sensor update rate (in the most extreme case temporarily disabling a particular sensor's updates) to influence the sensor reading
reaching the controller.
We show that this reconfiguration can be achieved through a single malicious message -- in contrast to established sensor spoofing attacks, where the intruder continuously injects malicious traffic. Thus, the \attackShort requires only intermittent access and limited resources and can persist after the attacker is out of range.
In our proposed attack, the CPS destabilization is caused by the reconfigured sensors, without otherwise manipulating code execution, sensor measurements, or denying communication.

We first investigate practical realizations of these abstract attacks on Commercial-Off-The-Shelf (COTS) flight controllers.
The attack leverages the fact that sensor chips allow the configuration via the sensor's serial APIs.

Then we formalize the attacks and evaluate the system security implications of SDA at a hardware, RTOS, control, and anomaly detection level
Finally, we propose a new automated attack synthesis algorithm using reinforcement learning to allow malicious drone control, instead of simply crashing.

Our results show that \attacksShort induce severe consequences on UAV stability, quickly leading the drone to deviate from the planned trajectory or crash. Our proposed \attacksShort impacts the controller performance without actively altering it (i.e.,~without causing any abnormal operation into the flight controller code).
Moreover, our results show that our novel attack by exploiting the sensor reconfiguration (instead of continuous spoofing~\cite{jang23paralyzing}) remains stealthy from state-of-the-art Anomaly Detector~\cite{choi18CI,khan2021m2mon}. Our takeaway is that, although sensors are fundamental for enabling autonomy in current UAV architectures, their insecure configuration poses a significant trust issue,  affecting the whole system's security and functionality. Finally, we discuss potential countermeasures.

The contributions of the paper are:
\begin{itemize}[noitemsep]

\item We propose a novel approach in which the attacker indirectly manipulates sensor readings via stealthy reconfiguration of the sensor via a shared bus. We show that this reconfiguration (e.g.,~sampling rate changes or suspensions) can lead to unexpected results on the reported sensor values, and are transparent to the controller.
\item We investigate the effects of such manipulations on a real-world UAV controller platform (i.e.,~ArduCopter), and show that our manipulations can even impact control loop timing to stall control.

\item We experimentally implement the three stages of the attack in different environments and demonstrate general feasibility. We show the impact of reconfigurations on five different hardware platforms and identify severe effects on two different control software stacks.

\item We propose an attack synthesis methodology to optimize \attacksShort for deterministic manipulation of drone behavior. This is the first time that drone manipulations do not only lead to uncontrolled crashes but control of the drone without active sensor spoofing.

\end{itemize}

\section{UAV Security: Background and Motivation}
\label{sec:background}
\begin{figure}[tb]
    \centering
    \includegraphics[width=\columnwidth]{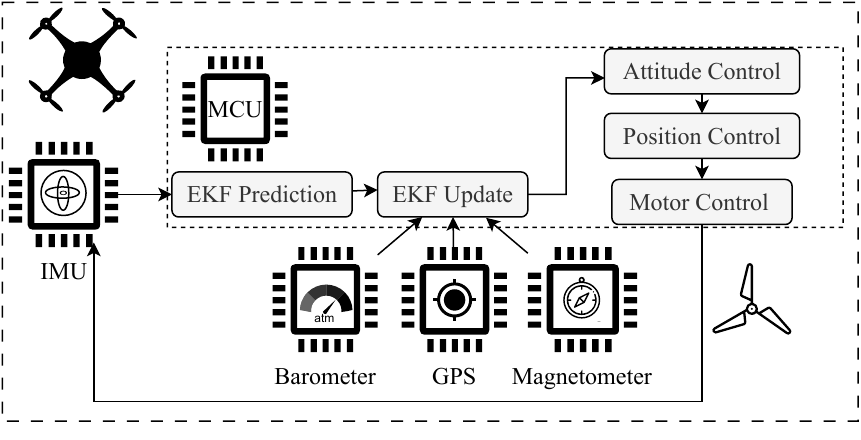}
    \caption{Flight control---sensors, GPS receivers and their contribution to the state estimation.}
    \label{fig:system_architecture}
\end{figure}
\subsection{Architecture of UAVs}
\label{background:architecture}
Autonomous UAVs are composed of sensors and actuators connected to a flight controller (Figure~\ref{fig:system_architecture}). Sensors observe the UAV state and report it to the flight controller. The flight controller is equipped with a microcontroller unit which is used to run the control and communication software. The received sensor readings are then used for state estimation via sensor fusion algorithms (e.g.,~Kalman Filter). Based on the estimated state and the planned trajectory, the flight controller computes the control action, which is finally transmitted to the actuators (motors with propellers).

\Par{Sensors for autonomous navigation} Inertial Measurement Unit (IMU) is an essential sensor for flight stabilization. It uses an accelerometer and gyroscope to report the proper acceleration and the angular velocity to the flight controller, which utilizes this information for attitude estimation (estimating the pose of the drone). Other sensors required for autonomous navigation are the compass, the magnetometer, and the barometer. Moreover, a GPS receiver is necessary to locate the drone. Readings from all these sensors and the GPS receiver are input to the Kalman Filter for state estimation.

\Par{EKF architecture} Extended Kalman Filter (EKF) is the non-linear extension of the Kalman Filter~\cite{kalman1960new}, it operates in two steps, prediction and update.  During the prediction step, the state estimate is computed (e.g.,~roll, pitch, yaw, velocity), while during the update the predicted state is improved by fusion with other sources.

Popular open-source flight control software such as Ardupilot applies state-of-the-art Extended Kalman Filter state estimation for autonomous navigation.
According to the documentation~\cite{kalmanfilterardupilot}, state prediction relies on IMU data, while the state update is done by fusing the other available sensor readings.
State prediction is calculated every time new IMU data is received (IMU has a high sampling rate), while state update is performed less frequently (as the other sensors are sampled less often). The IMU sensor reading is the most important and has the highest priority in the state estimation procedure. Given its pivotal importance in the state estimation, we concentrate on IMUs as target sensors.

\subsection{Serial Buses for Sensor Communication}
\label{sec:serial_busses}

In the UAV architecture, sensors communicate via serial buses to the microcontroller. Examples of these serial buses are SPI, I2C, CAN, and UART. These serial bus protocols are not authenticated or encrypted. Such protocols are used for sensor readings transmission, and for sensor configuration, done by the microcontroller via serial APIs. Those serial APIs are described in the manufacturers datasheet, publicly available online. The application running on the microcontroller changes the configuration registers on the sensor to configure the sensors to satisfy the application's real-time control requirements.

For this paper, we introduce the readers to the I2C protocol, which is commonly used for serial device communication. The I2C protocol uses a two-line bus to communicate among devices, a clock signal, and a data line. All communication is coordinated by a central controller, typically the main MCU of the system. All other bus members listen to the data line and respond to the specific controllers' requests. The main controller mandates when the devices communicate via the clock signal.

\subsection{Attacks on UAVs}
\label{sec:priorattacks}
Nassi et al.~\cite{nassi2021sok} systematized the state-of-the-art UAVs' attacks and defenses. We summarize some of the attacks investigated by prior works. Son et al.~\cite{son15rocking} demonstrated how sound injection attacks can affect gyroscopic sensors and affect drone controllability. While Jeong et al.~\cite{jeong23unrocking} investigated how to remediate sound injection attacks. Sathaye et al.~\cite{Sathaye2022GPS} experimentally investigated sensor spoofing attacks on drones and demonstrated that the takeover and control of a drone are challenging and require real-time spoofing signal manipulation. Dayanıklı et al.~\cite{dayanikli22physical}, investigated the use of false PWM actuation commands via IEMI to adversarially fly a victim UAV plane. Jang et al.~\cite{jang23paralyzing} investigated the effect of IMU signal corruption by actively blocking serial communication channels by utilizing intentional electromagnetic emission investigations to make the sensor readings at the controller noisy and crash a UAV. Selvaraj et al.~\cite{selvaraj2018electromagnetic}, demonstrated how via IEMI, analog sensor transmission can be manipulated to produce arbitrary sensor and actuator values, moreover, they show the induction of bit flips in serial communications. Dayanıklı et al.~\cite{dayanikli2022wireless}, demonstrated bidirectional bit flips via IEMI in serial protocols such as UART and I2C. Zhang et al.~\cite{zhang2023electromagnetic}, demonstrated the effects of IEMI on differential signals, showing that such systems are also vulnerable to IEMI induction. Xie et al.~\cite{xie2023bitdance}, demonstrated UART bit-level bus manipulation via IEMI injection and proposed a countermeasure to mitigate such attacks.

\subsection{Anomaly Detection for UAVs}
\label{sec:priordetectors}
Open-source autopilots such as Ardupilot deploy the EKF failsafe~\cite{EKFfailsafeardupilot} mechanism, it monitors the variance of the Extended Kalman filter innovation (residuals) and applies threshold-based triggers to classify an anomaly on the drone. According to the specification, the EKF failsafe is triggered when the EKF innovation exceeds the threshold for more than $1$ second. As a detection response, the drone will switch to land mode. Similarly, prior work in the security for Robotic Vehicles proposed in-vehicle anomaly detection. Quinonez et al.~\cite{quinonez20savior} proposed to detect anomalies by applying an additional Extended Kalman Filter monitor and the CUSUM change detection algorithm on the filter residuals. Choi et al.~\cite{choi18CI} proposed to use control invariants derived via system identification to detect anomalies in the target system. More recently, Khan et al.~\cite{khan2021m2mon} proposed to detect attacks affecting the control system by monitoring the microcontroller's low-level peripheral bus registers to identify any abnormal access patterns, and by moving the Kalman filter estimation outside the real-time operating system (RTOS).

\begin{table}[]
    \centering
    \caption{Attacker model comparison. Compared to prior work attacks against UAV sensors, our attack does not require continuous signal injection, active sensor value spoofing, or sensor corruption. Instead, our attacker only requires injecting a single reconfiguration message on the bus, the manipulation is persistent even when an attacker is out of range.}
    \label{tab:attacker_comparison}
    \resizebox{\columnwidth}{!}{
    \begin{tabular}{r|ccccc}
    \toprule
        Spoofing&  Continuous & Active & Sensor &Single &Persistent\\
        Type&  injection & spoofing& corruption & injection& Out of Range\\
        \midrule
        GPS, SEC'22 \cite{Sathaye2022GPS} & \Tdot & \Tdot  & \TDot & \TDot&\TDot\\
        IEMI, NDSS'23 \cite{jang23paralyzing} & \Tdot & \TDot & \Tdot & \TDot & \TDot \\
        Acoustic, NDSS'15 \cite{son15rocking}&\Tdot & \TDot & \Tdot & \TDot & \TDot\\
        Acoustic, NDSS'23 \cite{jeong23unrocking} & \Tdot & \TDot & \Tdot&\TDot&\TDot\\
        Our SDA &\TDot & \TDot &\TDot & \Tdot&\Tdot\\
    \bottomrule
    \end{tabular}}
\end{table}

\subsection{Limitations in Existing Work, Research Gap and Impact} Current UAV architectures are vulnerable to remote sensor manipulations. Process-based anomaly detection systems~\cite{choi18CI,quinonez20savior,khan2021m2mon} can detect such manipulations, enabling recovery~\cite{dash2024diagnosis}.

Our proposed attack model poses new threats to drones in real-world applications (such as military operations or consumer applications) by circumventing anomaly detection (and consequently making recovery ineffective). Table~\ref{tab:attacker_comparison} summarizes the differences between our novel \attacksShort attack and prior proposed sensor spoofing or false data injection attacks (see Section~\ref{sec:priorattacks}). In contrast to prior work that continuously manipulates the sensor readings~\cite{jang23paralyzing,Sathaye2022GPS,son15rocking,jeong23unrocking}, we propose to reconfigure the sensor permanently, as a result our attack persists after the attacker is out of range.
The traffic leading to control destabilization is generated from the legitimate source without breaking the code execution or corrupting the communication channel. Moreover, unlike existing IMU sensor attacks that primarily aim to crash the drone, our SDA can direct the drone to a specific target location.

\Par{Scope of this Work} This work explores the systems security implications of the proposed \attacksShort across the UAV hardware and software architecture. While previous research has demonstrated UAV sensor interference via IEMI to destabilize the vehicle, our attack leverages IEMI interference as a possible way to trigger sensor reconfiguration. The IEMI aspect is acknowledged as an established attack vector~\cite{dayanikli2022wireless, dayanikli22physical, jang23paralyzing,xie2023bitdance,zhang2023electromagnetic}, we do not claim it as a novel contribution.

\section{System model and Assumptions}
\label{sec:system}

\subsection{System Model}

We consider a UAV equipped with a flight controller (Figure~\ref{fig:system_architecture}) running state-of-the-art autopilot software, as described in Section~\ref{background:architecture}.

A monitoring system like M2MON~\cite{khan2021m2mon} is in place to detect attacks on the UAV by monitoring the Kalman filter and the serial register. In its most basic form, the monitor will check for unexpected large differences in consecutive sensor readings (e.g., bad data detection~\cite{cardenas2011attacks}). If an attack is detected, the drone will switch to failsafe mode or execute the recovery~\cite{dash2024diagnosis}.

\subsection{Threat Model}
An attacker is assumed to be able to access white-box to a copy of the victim's hardware (e.g.,~by buying the same UAV drone or flight controller).

The attacker's goal is to deviate the victim drone from its expected trajectory or crash it, without being detected by the monitoring system. In order to achieve this goal, the attacker aims to carefully deny or delay specific sensor readings from reaching the controller running on the MCU.

The attacker can inject energy into the serial communication channels (e.g., I2C bus) on the drone. This can be achieved in various ways: i) a supply chain attack, implanting malicious behavior into one component connected to the bus~\cite{zhang2024control}, ii) by remotely injecting (or changing existing) messages on the bus via IEMI~\cite{dayanikli2022wireless,dayanikli22physical}.

The attacker does not have remote code execution on the main MCU since executing attacker code there can also be assumed to trigger detection measures such as M2MON.

\subsection{Research Goal and Challenges}

In a nutshell, the idea of the attack is twofold: i) the attacker

reconfigures sensors (e.g. their measurement frequency) and thus influence sensor measurements stealthily, and ii) using this capability, the attacker forces the victim controller to perform wrong control decisions, leading the drone on the wrong path or crash.
To capture the intuition of this novel generalized type of attack targeting configuration of sensors, we introduce the term \emph{\attacks}.

There are two main challenges: \textbf{C1)}~It is not clear from prior work which manipulations are possible for the attacker in this scenario without raising errors or alarms at the controller and \textbf{C2)}~Depending on the impact the attacker can have on the sensors, deterministically controlling the effects on the drone behavior will be challenging. We claim that attacks on cyber-physical systems in which the attacker denies timely sensor updates represent an interesting class of under-explored attacks in three main aspects: a)~What are options for the attacker to lead to the \attacks,  b)~What are the effects of such \attacks on the input provided to the victim controller, c) What are the effects on the control decisions taken by the victims and how can the attacker make deliberate (as opposed to random DoS) perturbations on those decisions?

\section{\attacks}

In this section, we introduce \attacksShort and their system-level (sensor, controller, and actuation) effects. In Section~\ref{sec:hardware}, we present the attack vector (configuration messages by attacker),  providing examples of how it can be practically launched by an attacker, and discussing a re-configuration that alters the system behavior. In Section~\ref{sec:SDA}, we provide a formal attack definition by modeling the interaction of the attack with the dynamical control system (a drone in our case). In Section~\ref{sec:controllability}, we present the impact of the physical deviation of the victim's vehicle.

\subsection{Part 1: Sensor Reconfiguration}
\label{sec:hardware}
Commercially available sensors used in UAVs are configured by issuing (not cryptographically secured) bus write commands into specific register addresses (serial API, Section~\ref{sec:serial_busses}). All the commercially available sensors that we reviewed (BOSCH, TDK, STMicroelectronics, Analog Devices) allow configuration via serial APIs, and to the best of our knowledge, configuration commands for a sensor cannot be flushed to avoid sensor configuration overriding.
An attacker who alters the configuration of the sensor will influence the state estimation in the drone, leading to incorrect actuation. We demonstrate that modern controllers rely on the implicit trust of sensors.

During the normal operations of a drone, the sensor is periodically queried from the microcontroller---with a frequency of \tild1600Hz---to pull the fresh data (see Section~\ref{sec:serial_busses}). The attacker leverages such periodic messages or creates new traffic on the bus to achieve target senor reconfiguration. In the evaluation section, we provide two case studies showing how the attack can be practically launched on a victim's vehicle (locally via a hardware supply chain attack or remotely via IEMI).

\subsubsection{Target reconfiguration}

To achieve the intended effect on the target vehicle we identify two main options for a target reconfiguration, the first is the sensor suspension mode, while the second is the sampling frequency reduction.

a) Sensors allow suspension (also referred to as power-save mode, depending on the manufacturer) for power-saving purposes. In this mode, the sensor stops updating the sensor reading register. The control software will continue its normal execution but the information received from the external sensor will be no more reliable. It will be outdated, wrong, or absent depending on the manufacturer's specified behavior and the controller's behavior.

b) Sensor operating frequency can be modified. Modern flight controllers (e.g.,~the Pixhawk 6c~\cite{pixhawk6c}) rely on general-purpose IMUs that can be programmed for different applications (i.e.,~the same IMU chip can be used in various contexts, for example, in smartphones, activity trackers, etc.). In flight control software, sensors are configured to run at the highest sampling frequency allowed by the sensor to allows precise control of the aircraft. By reducing the sampling frequency, an attacker can delay sensor updates and compromise control accuracy. This can be modeled as intermittent suspension with a duration inversely proportional to the new frequency (For space constraints we discuss the sensor frequency reconfiguration and evaluate its effects at the system and control level in Appendix~\ref{sec:frequencyAttack}).

\begin{figure*}[t]
    \centering
    \includegraphics[width=0.7\linewidth]{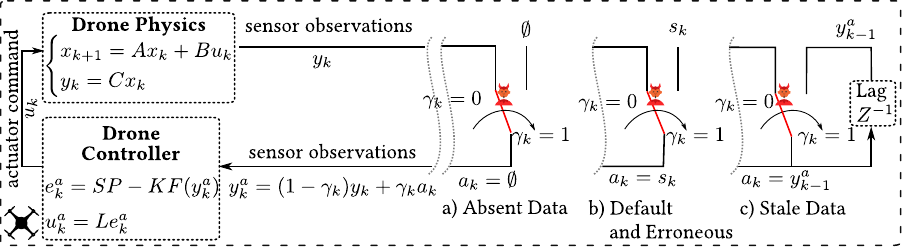}
    \caption{\attack abstraction. An attacker reconfigures sensors to manipulate the readings. This can have different effects in the control loop, a) Absent data, where no data is received b) Default data where $s_k$ is constant and Erroneous data where $s_k\sim~\mathcal{N}~(\mu,\,\sigma^{2})$, c) Stale data where last observation is received by the controller.}
    \label{fig:stale_attack_diagram}
\end{figure*}

\subsection{Part 2: Control Manipulation}
\label{sec:SDA}
In this section, we characterize the possible behaviors induced by the sensor reconfiguration at the interaction between the sensor communication and the controller. Moreover, we formalize the attack using control theory notation.

\subsubsection{Possible Behaviors Induced by \attacksShort}

Consider the system depicted in Figure~\ref{fig:stale_attack_diagram}. Based on the sensor readings, the controller computes the actuator signal to control the physical process. Intuitively, our \attacksShort are the attacker's actions to `interrupt' the feedback connection from the physical process to the controller (in Figure~\ref{fig:stale_attack_diagram}, the `switch' symbol corresponds to the attacker's action). After the attacker influences the system, an attacked set of observations reaches the controller.

In control theory, prior work proposed control strategies for lossy control packets~\cite{schenato2007foundations,Schenato2009tozerotohold,sinopoli2004kalman}, by stochastically modeling missing observations. In contrast, we assume the attacker deterministically triggers the \attacksShort communication to destabilize the process. We identify four possible behaviors resulting from an \attackShort, and their resulting missing observations (i.e., missing sensor readings). In Section~\ref{sec:IMUsExperiments}, we verify which of the hypothesized behavior occur in practice in COTS IMU sensors and verify their impact on the system control.

\begin{itemize}[noitemsep]
\item\Par{Absent data} No sensor values are received at the controller. This situation might occur when the sensor is disabled.

\item \Par{Default value data} The value that reaches the controller is a constant. This situation might occur when the sensor is in an error state and replies with a default value to the queries.

\item\Par{Erroneous data}  The value that reaches the controller follows a certain statistical distribution (noise). This might occur when some disturbances in the sensor prevent correct sensing (e.g., acoustic signal interfering with the sensor~\cite{son15rocking}) or disturbances in the communication channel~\cite{jang23paralyzing}.

\item \Par{Stale data} The sensor stops updating, and the last observation is retransmitted to the controller. Krotofil et al. introduced this setup in an ICS scenario~\cite{krotofil2014cps}.

\end{itemize}

Due to the influence of \attackShort on the sensor readings, the controller reacts depending on the received observations. In the case of \emph{absent data} two control strategies were proposed by Schenato et al.~\cite{Schenato2009tozerotohold}. \emph{Zero-input.} Stop actuating the system. \emph{Hold-input.} Keep actuating based on the last observed value (or keep the last control action). In the case of \emph{default data}, \emph{erroneous data}, and \emph{stale data} from the sensor, the controller will continue actuating the system based on the received data.

The SDA influences the state estimation at the Kalman filter, this will propagate to the controller producing the wrong control.

\subsubsection{Attack Formalization}
\label{sec:attack_formalization}
In  Figure~\ref{fig:stale_attack_diagram} we show the attack formalization. We consider a linear discrete-time system ($k$), Equation~\ref{eq:linear_system}.
\begin{equation}
\begin{cases}
    x_{k+1} = Ax_k + Bu_k\\
    y_k = Cx_k
\end{cases}
\label{eq:linear_system}
\end{equation}
where $A$, $B$, and $C$ are matrices that describe the coefficient physical model. $x_k$ denotes the current system's state, $y_k$ are the sensor readings, and $u_k$ denotes the actuation command that is computed based on the current sensor readings. In the case of \attacksShort, the observations that reach the controller are affected by the attacker ($y^a_k$) which can decide to start the attack on the system ($\gamma_k \in \{0,1\}$).

In the case of Absent data no sensor data is received at the controller during an attack, and the sensor observation model is

\begin{equation}
    y^a_k = (1-\gamma_k)y_k  + \gamma_k\emptyset
\end{equation}
In the case of Default value data, the observation model is

\begin{equation}
    y^a_k = (1-\gamma_k)y_k  + \gamma_ks_k
\end{equation}

 where  $s_k$ is constant, while for erroneous data  $s_k\sim \mathcal{N}(\mu,\,\sigma^{2})$. Finally, in the case of stale data, the observation model is
 \begin{equation}
     y^a_k = (1-\gamma_k)y_k  + \gamma_ky_{k-1}^a
 \end{equation} where  $y_{k-1}^a$ is the last sampled value by the sensor the attack.

Before reaching the controller (e.g.,~PID controller), $y^a_k$ goes through the Attitude and Heading Reference System (AHRS) and Kalman filter (where it gets fused with the other sensor sources to refine the system state and the sensor readings estimate, e.g.,~yaw, pitch, and roll) (see Figure~\ref{fig:system_architecture}). The attack-induced sensor reading propagates through the filters to reach the controller, where the wrong state estimate influences the error value \emph{$e^a_k$} (between the observation and the control target). Formally,
\begin{equation}
\begin{aligned}
\label{eq:error_to_actuator}
    & u^a_k = Le^a_k\\
    \text{where}\quad & e^a_k = SP - KF(y^a_k)\\
    \quad & y^a_k \in {\{y_k,\text{absent, default, erroneous, stale}\}}\\
\end{aligned}
\end{equation}
L refers to the controller parameters that operate based on the error, SP is the set point (target) for the controller and KF is the Kalman filter.

\subsection{Part 3: Physical Deviation}
\label{sec:controllability}

\begin{figure}[tb]
  \centering
  \includegraphics[width=\linewidth]{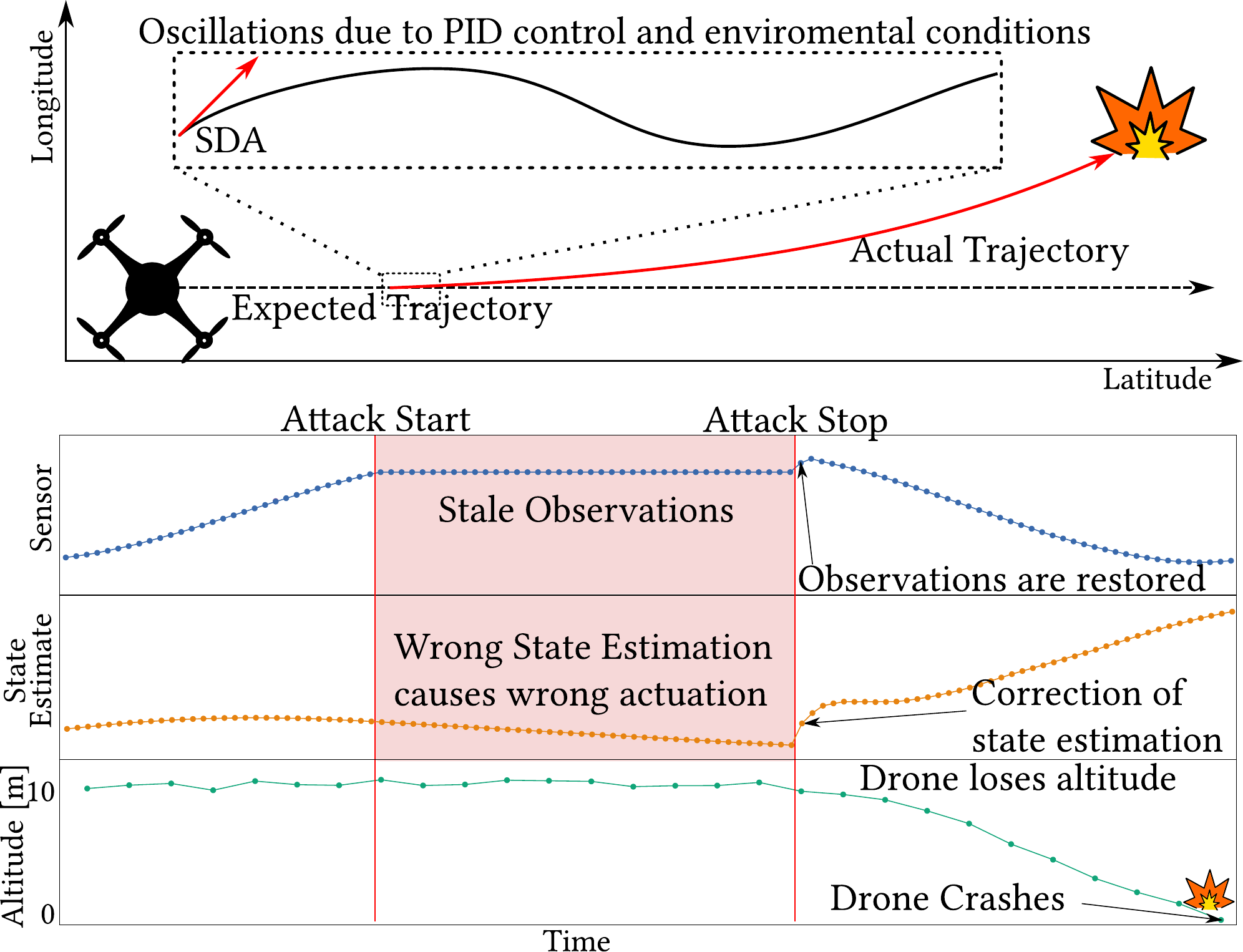}
  \caption{Motivating example, a drone flies on a straight trajectory. The flight is influenced by sensor and environmental noise which results in continuous correction applied to the flight. An attacker starts a \attack, targeting one of the drone sensors. The sensor stops updating the transmitted value to the MCU. Consequently, the attitude estimation is compromised, causing wrong actuation commands which results in the drone deviation or crash.}
  \label{fig:StaleAttack}
\end{figure}

An \attackShort induces the state estimation to be incorrect or delayed. This causes the wrong actuation in the system. In real-world control systems, the effect of measurement and environmental noises makes the PID controllers perform an error correction w.r.t. a target value (as explained in Equation~\ref{eq:error_to_actuator}). We are interested in understanding if this wrong actuation can be adversarially exploited to deviate the drone to an adversary's desired location.

Figure~\ref{fig:StaleAttack} shows an example: while a drone is following a specific mission (e.g. moving along waypoints), the attacker is launching targeted attacks to suppress specific (or all) new sensor readings (e.g.~through sensor reconfiguration).
When the \attackShort starts, the last computed attitude is the last correct estimate of the drone pose. The actuation command performed will reflect the correction w.r.t. the expected trajectory of the drone. Such attacks cause the drone controller to operate on outdated sensor readings with constant actuation -- which causes the drone to go on a wrong trajectory.
When the attack stops, the drone will compute the correct state estimation and attempt to re-stabilize the body frame. In Figure~\ref{fig:StaleAttack}, the drone crashes on the ground while recovering from the \attackShort.

In contrast with a sensor spoofing attack, where the attacker can arbitrary substitute the sensor readings to induce the desired behavior, in an \attacksShort the attacker does not actively decide the value received at the controller, but instead it would follow one of the four behavior described in the previous section (absent, default, erroneous, stale data).

To deviate the victim vehicle towards the adversary desired position the attacker intermittently activates and deactivates the \attacksShort to influence the sequence of actuation's ($u_k$) to minimize the distance between the drone location and adversary desired position which we refer to as adversarial control.

\begin{equation}
    \begin{aligned}
        \text{minimize} \quad & \sum_{k=0}^{t} {\sqrt{(\text{DronePos}_k - \text{AttackerGoalPos}_k)^2}} \\
        \text{where} \quad & \text{DronePos}_{k+1} = \sum_{k=0}^{t}{A(\text{DronePos})_k+Bu^a_k}\\
    \end{aligned}
    \label{eq:adversarial_control}
\end{equation}

In Section~\ref{sec:synthesis}, we propose an attack synthesis methodology for the adversarial control problem.

\section{Experimental Evaluation}

We run several experiments to comprehensively study \attacks. The first part involves the realization of \attacksShort in hardware, followed by how they affect the dynamical system managed by the flight controllers, and finally, a general perspective of UAV's controllability by studying the impact of such attacks on real-world flight controllers.

\subsection{Case Study: \attacksShort in Hardware}
\label{apx:alternativeattacker}
To study the \attacksShort realizations in Hardware, we design a set of experiments that help us to answer the following research question. \label{rq1}\textbf{RQ1}: \emph{How can the \attacks be practically launched on modern flight controllers?}
In this section, we demonstrate how an attacker can practically launch the proposed \attacksShort on hardware. We offer two case studies to showcase the attack.

\subsubsection{Local sensor reconfiguration via unauthorized device}\hfill

\Par{Attack Setup}  The attacker is implemented as a third-party peripheral (Raspberry Pi) connected to the I2C bus. The main controller continuously reads data from the IMU. I2C uses pull-up resistors in both lines to ensure a logical `1', which means only `1' can be turned into `0' and not vice-versa. Making message manipulation challenging, as the attacker lacks full control of the data line.

We found that the clock signal is generated only when the main controller sends a request on the bus. This presents an opportunity for an attacker to inject malicious commands. When the bus is idle, any device could generate the clock signal and maliciously impersonate the legitimate controller.
By listening to the reading patterns on the shared bus, the attacker can determine when to inject crafted commands, avoiding collisions. The attacker-controlled peripheral performs the injection by unilaterally reconfiguring its I2C interface from peripheral to controller mode.

\begin{figure}
    \centering
    \includegraphics[width=0.8\columnwidth]{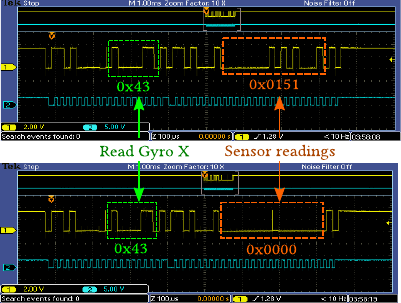}
    \caption{Reading Gyro X on the shared bus. Top: Readings before the malicious command injection. Bottom: Readings after the malicious command injection, no data are transmitted due to the device reconfiguration.}
    \label{fig:toy_experiment}
\end{figure}

\Par{Injection result} The attacker has a time window before the next communication with the legitimate controller occurs (legitimate communications occur at 400Hz in Ardupilot). Within this time window, our attacker sends requests to the target peripheral on the bus to change its configuration. After injection, the sensor stops answering the controller requests. For example, before the attack, the read gyro X command \texttt{0x43} is followed by a sensor reading response \texttt{0x0151}, and after the attack the sensor responds \texttt{0x00}.
Figure~\ref{fig:toy_experiment}, shows the impact of the impact of the \attackShort on the sensor. After reconfiguration, the sensor will stop answering requests from the MCU, demonstrating  the attack practicality.

\begin{figure}
    \centering
    \includegraphics[width=0.35\columnwidth]
    {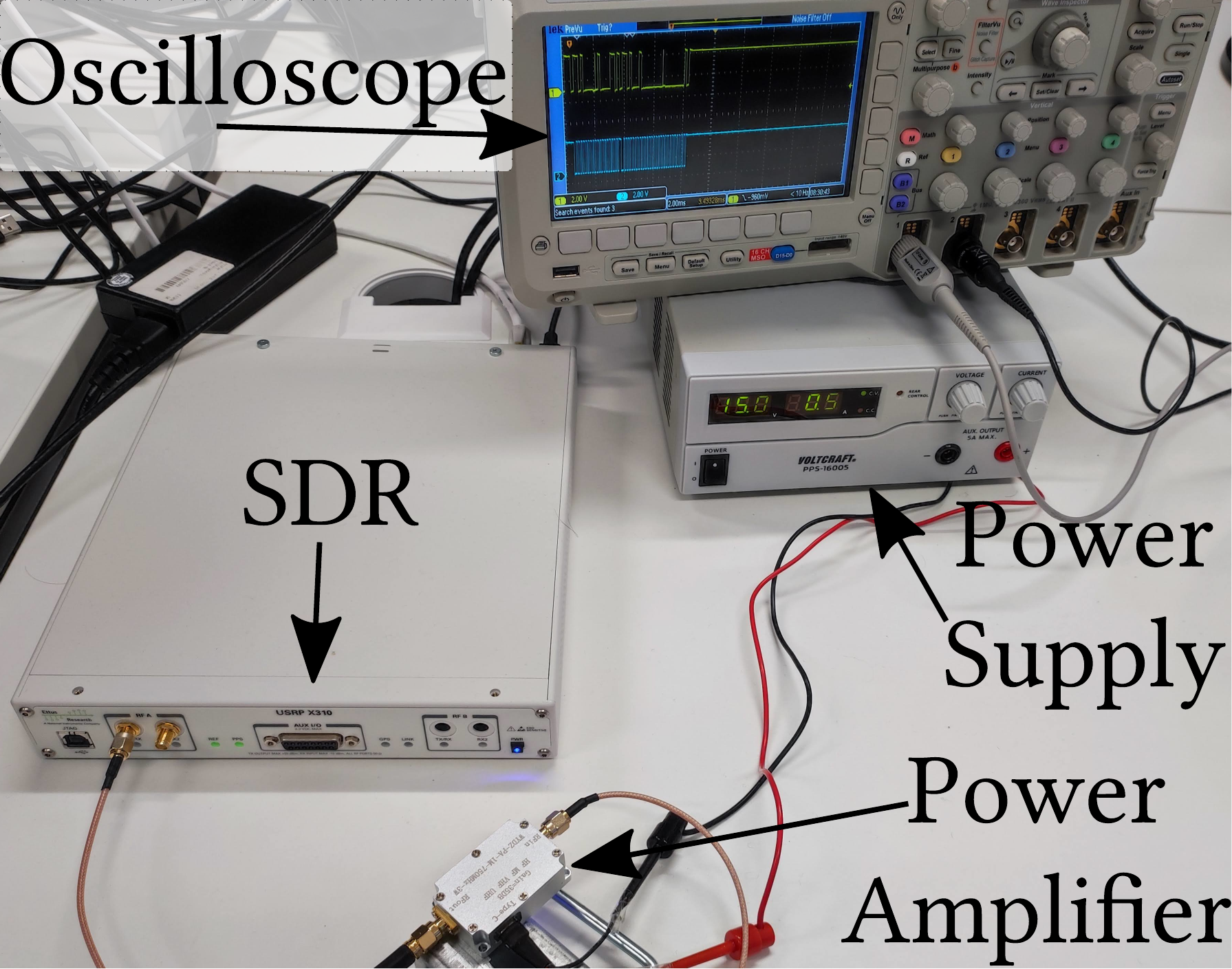}
    \includegraphics[width=0.49\columnwidth]{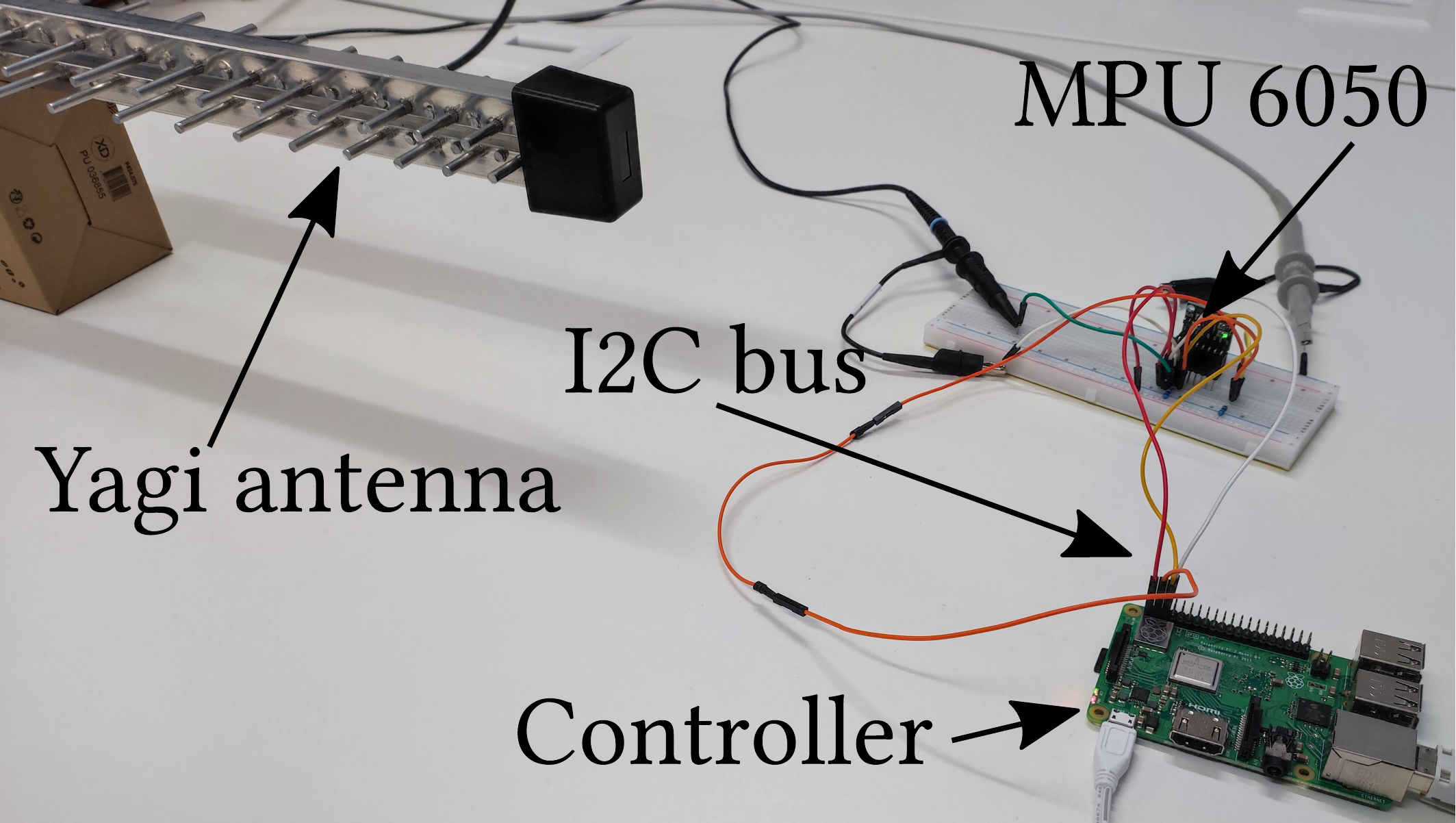}
    \caption{IEMI injection testbed. On the left: the SDR, the power amplifier (connected to the power supply), and the oscilloscope. On the right, the Yagi antenna, the Raspberry connected to the IMU.}
    \label{fig:testbed}
\end{figure}
\subsubsection{Remote sensor reconfiguration via IEMI}\hfill

\Par{Attack Setup}  The attacker is injecting wireless electromagnetic emission on the victim bus. Prior work has shown that random values can be injected in the shared bus via IEMI~\cite{jang23paralyzing}. In this section, we demonstrate that in principle an attacker can use IEMI to perform a targeted sensor reconfiguration on a victim's sensor.
Consistent with prior work~\cite{jang23paralyzing}, we set up the same testbed with a main controller (Raspberry Pi3) that connects to an IMU (MPU6050) through a shared bus (I2C). A Python script that relies on the \verb|smbus2| library controls the I2C bus communication.

\begin{figure}[tb]
  \centering

      \includegraphics[width=\linewidth]{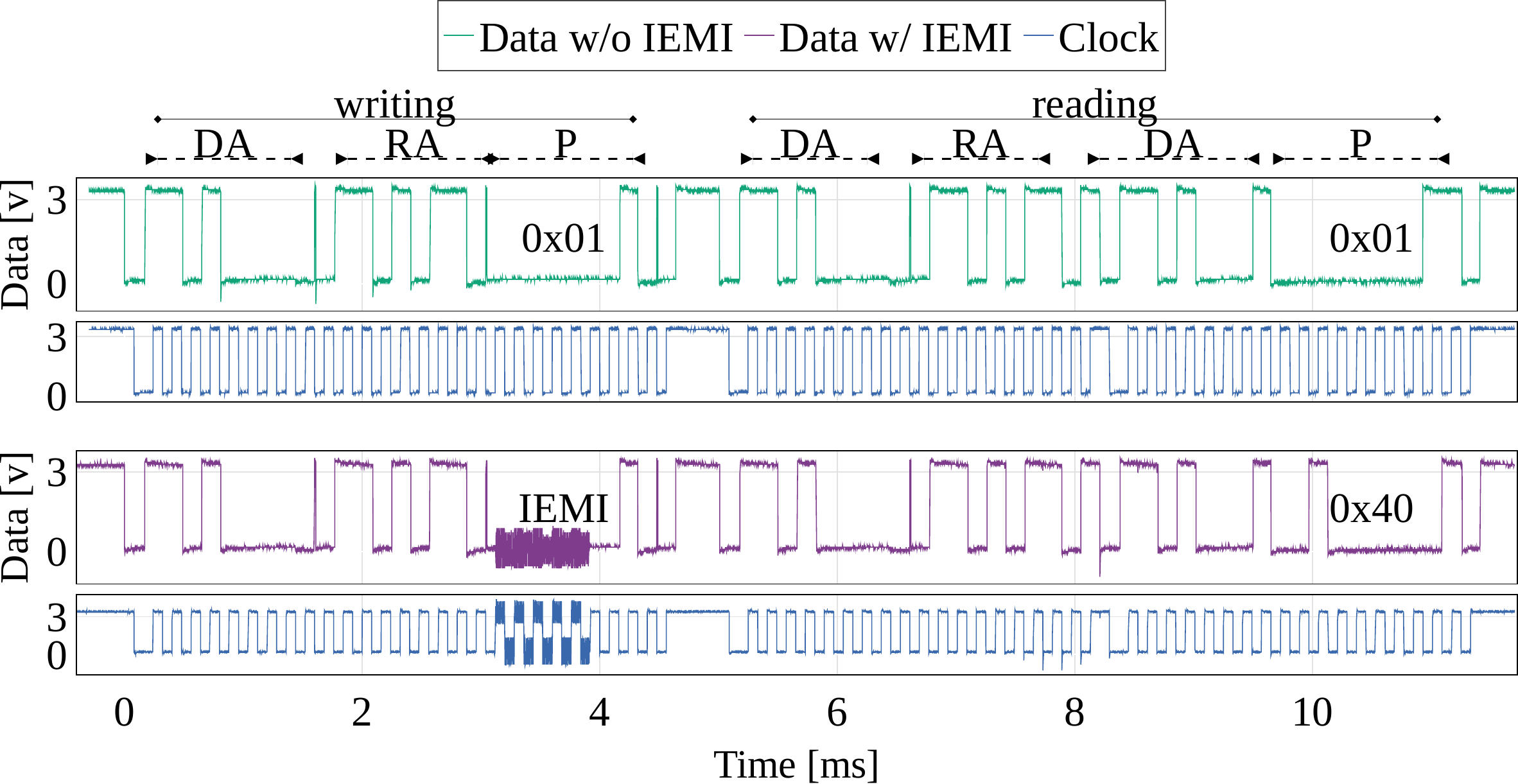}

    \caption{I2C bus oscilloscope capture. Comparison of stored register value before and after triggering suspend mode in the sensor via IEMI attack on MPU6050. The IEMI modifies the payload of the message from 0x01 to 0x40.
      DA: Device address, RA: Register address, P: Payload.}\label{fig:MPU6050attacks}

\end{figure}

For IEMI injection, we use a software-defined radio (SDR, USRP X310) connected to a power amplifier (WYDZ-PA-1M-750MHz) that provides power to a Yagi antenna (see Appendix Figure~\ref{fig:testbed}).
We monitor the IEMI effects on an oscilloscope (Tektronix MSO2024B) connected to the data and clock lines of the I2C bus. The I2C controller (Raspberry Pi3) controls how the sensor operates by changing the values on particular registers. To reconfigure a sensor, an attacker needs to inject a valid writing command in the shared bus. A valid I2C writing command is composed of a Device Address, followed by a Register address and a Payload (see Figure~\ref{fig:MPU6050attacks}). To achieve their goal, the attacker can bit-flip parts of register address or payload in write command to induce the desired misconfiguration.

To perform a successful IEMI injection, we must find the correct frequency that induces enough power to cause bit flipping in the serial bus.

After performing a frequency sweep, we found that the range between 120MHz and 150MHz injects enough power to disrupt the bus communication. We use 126MHz as the carrier frequency which showed the highest injection power.

As a proof of concept, we demonstrate the change of payload value on the fly. The attacker could potentially also perturb other message fields in the bus.

\Par{Injection result} Our goal is to achieve the proposed sensor suspension and sampling frequency reconfiguration via IEMI. According to the sensor specification~\cite{mpu6050}, to achieve the sensor suspension in the MPU6050, we have to write in the \verb|Power Management 1| register \verb|0x6B|, the value \verb|0x40|. Figure~\ref{fig:MPU6050attacks}, represents the collected power trace at the oscilloscope, with and without injection. The collected power traces contain a register write followed by a register read. As we can observe when the IEMI injection is performed (lasting 800$\mu$s, i.e., 5 clock cycles), the response from the register changes, and the value that is stored in the register is \verb|0x40|, which is our target value for suspend mode.

\subsubsection{Generalizability to other protocols} We note that the same approach applies to other serial protocols (like CAN). Independently of the bus protocol used, two factors make the reconfiguration possible, i) the use of unauthenticated buses and ii) the reconfiguration feature on the target sensor.

\begin{formal}
     To answer \textbf{\hyperref[rq1]{RQ1}}, we practically realize and demonstrate the reconfiguration command injection on an I2C bus testbed. Our case studies show that it is possible to maliciously reconfigure the sensor (locally or remotely). The proposed reconfiguration approach brings advantages compared to direct spoofing of the sensor readings that could potentially lead to collisions on the bus, making the attack detectable.
\end{formal}

\begin{figure*}
  \centering
  \subcaptionbox{BMI055\label{fig:BMI055Suspend}}{\includegraphics[width=0.252\linewidth]{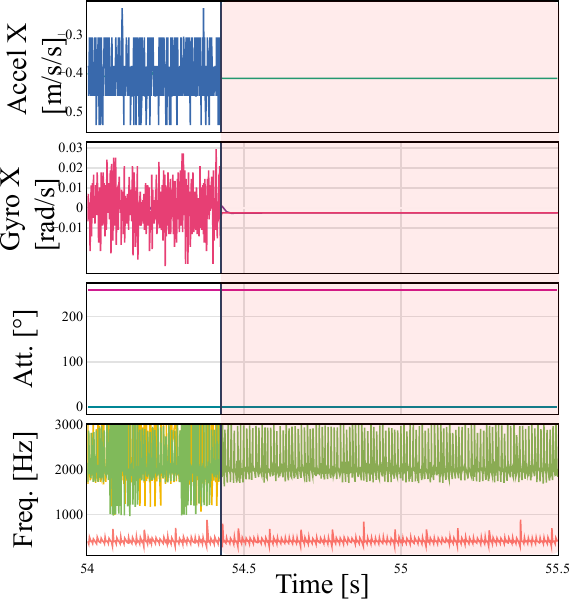}}
  \subcaptionbox{BMI270
  \label{fig:BMI270Suspend}}{\includegraphics[width=0.23\linewidth]{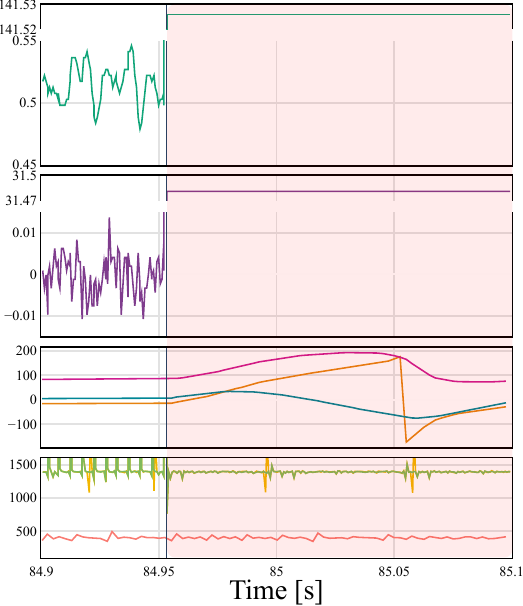}}
  \subcaptionbox{ICM42688\label{fig:ICM42688SSuspend}}{\includegraphics[width=0.1655\linewidth]{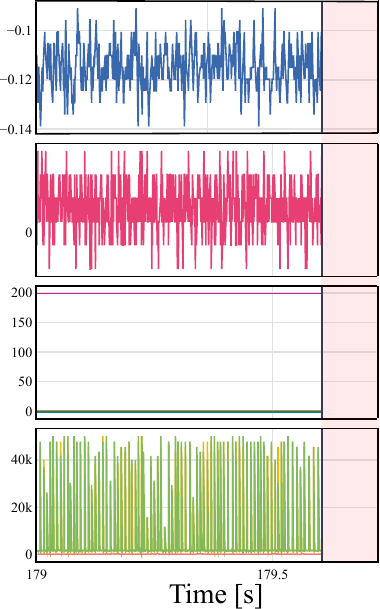}}
  \subcaptionbox{MPU6000\label{fig:MPU6000Suspend}}{\includegraphics[width=0.309\linewidth]{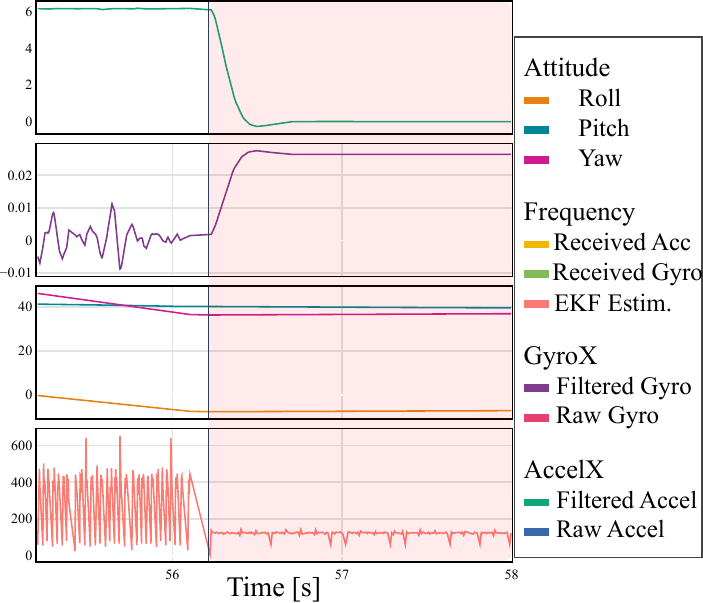}}
  \caption{Suspend attack behavior on the four tested IMUs. Red background: IMU suspended.}
\end{figure*}

\subsection{Control Manipulation: Reconfiguration Variations for Different Sensors}
\label{sec:IMUsExperiments}

Motivated by the practicality of reconfiguration attacks, we now investigate the reaction of IMUs to such reconfigurations, and the consequences on the flight control. In this section, we address \label{rq2}\textbf{RQ2}: \emph{Which practical interactions between an IMU and the control program are observed and what is their relation to the four hypothesized behaviors for \attacksShort?}

To verify the behavior of \attacksShort via reconfiguration attacks over the control algorithm, we run experiments on multiple COTS flight controllers and evaluate how different IMUs respond to the reconfiguration attacks. We also evaluate how the control software reacts to those reconfigurations.
We rely on the Ardupilot because it is a widely-used flight control software for commercial UAVs, offering state-of-the-art autonomous flight controllers. Additionally, for our evaluation of system-level effects of the \attacksShort, using either PX4 or Ardupilot would yield similar results since both rely on the same Extended Kalman Filter implementation~\cite{priseboroughInertialNav}.

\Par{Evaluation Setup} To test the IMU \attacksShort, we flash the flight controller with Ardupilot firmware, enable the logging, place the flight controller on a table, and launch the IMU reconfiguration during the flight control software execution. We then analyze the collected log files to assess the impact of sensor reconfiguration.

The flight controllers used in our evaluation are i) Kakute H7 Mini (BMI270
IMU); ii) Pixhawk 6C with a dual-IMU configuration (BMI055
IMU and ICM-42688-P IMU); iii) Omnibus F4 (MPU6000 IMU). All the considered IMU peripherals connect to the MCU via the SPI bus. Moreover, to assess the impact of the attacks in the real world, we evaluate the effects of the attacks on the Crazyflie V2, equipped with BMI088 IMU and the optical flow sensor.

\begin{table*}[]
    \centering
    \caption{Summary of IMU behaviors observed during Suspend mode attack}
    \resizebox{\textwidth}{!}{
    \begin{tabular}{lcccccccccccccc}
        \toprule
        & \multicolumn{7}{c}{Accelerometer Value} & \multicolumn{7}{c}{Gyroscope Value}\\
        \cmidrule(l){2-8}\cmidrule(l){9-15}
        & & \multicolumn{2}{c}{X} & \multicolumn{2}{c}{Y} & \multicolumn{2}{c}{Z} &  & \multicolumn{2}{c}{X} & \multicolumn{2}{c}{Y} & \multicolumn{2}{c}{Z}\\
         \cmidrule(l){3-4}\cmidrule(l){5-6}\cmidrule(l){7-8}\cmidrule(l){10-11}\cmidrule(l){12-13}\cmidrule(l){14-15}
        IMU Model    & Behavior&$\mu$&$\sigma$&$\mu$&$\sigma$&$\mu$&$\sigma$&Behavior&$\mu$&$\sigma$&$\mu$&$\sigma$&$\mu$&$\sigma$\\
          \cmidrule(l){1-1} \cmidrule(l){2-8}\cmidrule(l){9-15}
        BMI055   & Absent Data & - & - & - & -&- &- & Default & -0.0025  & 0.0  & -0.002 & 0.0 & 0.0012 & 0.0 \\
        BMI270   & Default & 141.53 & 0.0 & 141.53 & 0.0 & 141.53 & 0.0 & Default & 31.48 & 0.0 & 31.49 & 0.0 & 31.49 & 0.0\\
        ICM-42688-P  & Absent Data & - & - & - & -&- &- & Absent Data & - & - & - & -&- &- \\
        MPU6000  & Default & 0.2477 & 1.058 & 68.9905& 14.563& -0.3031 & 1.295 &Default & 0.0254 & 0.0 & -0.0346 & 0& 0.1211 & 0\\

        \bottomrule
    \end{tabular}
    }

    \label{tab:IMU_suspend_behaviors}
\end{table*}

\subsubsection{Effects of IMU Suspend}
For each IMU, we review the manufacturer's publicly available datasheet to identify the register/value pair used for device reconfiguration and verify its effect on the control software. We now detail our findings.

\Par{BMI055} IMU is composed of two sensor modules, the accelerometer and the gyroscope. The accelerometer reacts by stopping sending values to the MCU (Absent data), while the gyroscope reacts to the configuration change by reporting a constant value (Default Value data). Table~\ref{tab:IMU_suspend_behaviors}, reports the default value that was sensed on each axis during the attack, while  Figure~\ref{fig:BMI055Suspend} shows the results of the suspend reconfiguration on the BMI055 sensor. The Ardupilot controller reacts to the (partially) Absent data by applying the hold strategy on the last observed value, and the attitude estimation continues despite no accelerometer samples being received. As we can notice from the bottom plot in Figure~\ref{fig:BMI055Suspend}, the frequency of the EKF estimation is not affected by the reconfiguration attack.

\Par{BMI270} In this case, both the sensor modules react to the reconfiguration with the Default Value Data, and the reported value for each module is an out-of-range value (see Table~\ref{tab:IMU_suspend_behaviors}).  Figure~\ref{fig:BMI270Suspend}, shows the sensed behavior when starting the suspend attack on the BMI270. Ardupilot accepts this received value as legitimate and this will induce an error in the attitude estimation. The attitude estimator will believe that the drone is flipping (instead the drone is steady on a table). Also in this case the reconfiguration does not affect the sensor and EKF frequency, leaving the scheduler unaltered.

\Par{ICM-42688-P}
In this case, the IMU reacts to the suspend attack by stopping transmission of both the gyroscope and the accelerometer sensor updates. Since no data is received at the controller (for both sensors), no state estimation will be performed (see Figure~\ref{fig:ICM42688SSuspend}). This result shows the resource dependency between the IMU and the state estimation.  In Ardupilot, the state estimation is part of the \verb|fast_loop()| tasks, executed with the highest priority by the scheduler with a frequency of 400Hz. The control flow inside the flight controller remains unaltered, but IMU data's unavailability will make the highest priority task in Ardupilot not executed.

\Par{MPU6000} As reported in Table~\ref{tab:IMU_suspend_behaviors} the sensor behaves similarly to the BMI270 and will start reporting default values; those values are close to zero. In this case, after the reconfiguration, the sensor will also increase its responsiveness time (the attitude and EKF frequency drops from 400Hz to 100Hz, see  Figure~\ref{fig:MPU6000Suspend}),

\subsubsection{End-to-end Attack Demonstration}
We evaluate the impact of the reconfiguration attack strategies on the CrazyflieV2 platform equipped with the control firmware \verb|2023.02| and BMI088 IMU sensor. An attack demonstration video is available at~\textcolor{blue}{\url{https://youtu.be/Egnw2dHKqw0}}. The crazyflie device is hovering above the ground and it will continue doing so  (baseline in the video) unless an attack is started. In BMI088, the suspension produces absent data behavior (similar to BMI055). The impact of the suspended attack results in the drone crashing on the ground. Since the sensor updates are suspended, this attack is undetectable from a bad data detector (see Section~\ref{sec:detection}).

\begin{formal}
 To answer \textbf{\hyperref[rq2]{RQ2}} we characterize the behavior of IMUs when targeted with the proposed reconfiguration and their implications on the control algorithm (data unavailability blocks execution of tasks in the RTOS). Our results show that the proposed techniques will result in three main behaviors, namely, Absent, Stale, and Erroneous data (consistently with the behaviors hypothesized in Section~\ref{sec:SDA}).
\end{formal}

\subsection{Physical Deviation: Effect of \attacksShort at Control Level}
\label{sec:emulation}
In this section, we explore the impact of the proposed attacks on the dynamic system in an emulated environment. Specifically, we want to address the following research question
\label{rq3}\textbf{RQ3}:~\emph{What is the impact of the proposed \attacks on UAVs?}

\subsubsection{Evaluation Setup} For our evaluation, we assess the effect of \attacksShort on the Ardupilot~\cite{ardupilot} autopilot system (version 4.3.3), running in the Software In The Loop (SITL) simulator. SITL is a fully functional environment to run Ardupilot software in simulated physical surroundings. All the Ardupilot features are available within the SITL framework, for this reason, it is well suited to test the impact of \attacksShort on the control algorithms without safety hazards. To emulate the effects of the \attacksShort we target the virtual sensor driver for the IMU (\verb|AP_inertial_sensor_SITL.cpp|). In turn, we emulate the Stale data by storing the last observation and re-transmitting it to the attitude controller, the Default data by transmitting a constant value, the Erroneous data by transmitting random data, and the Absent data by skipping the data transmission. We create a straight line mission, with a desired altitude of 50 meters and fly it $55$ times, $11$ times without attacks to create a baseline trajectory, and 11 times launching each \attackShort for the duration of $1$ second to evaluate the impact of the attack. The \attacksShort is launched when the drone reached the desired altitude and follows the straight trajectory. We evaluate the impact of the attacks in terms of deviation $D$ (Equation~\ref{eq:deviation}) as the instantaneous mean squared difference between the attack trajectory $A$ and the reference trajectory $R$. The distance is computed over the three-dimensional space axis, x (longitude displacement), y (latitude displacement), and z (altitude displacement).
\begin{equation}
    D_{t(x,y,z)} =  \sqrt{(A_{t(x,y,z)}-R_{t(x,y,z)})^2}
\label{eq:deviation}
\end{equation}
In particular, we consider Maximum deviations Equation~\ref{eq:max} induced by the attack (i.e.,~between the events \emph{attack start} and the \emph{crash}, \emph{failsafe} or \emph{landing} event timestamps).
\begin{equation}
    \text{max}  \quad  D_{t(x,y,z)}  \quad  t \in {\text{\{attack start, failsafe/crash\}}}
    \label{eq:max}
\end{equation}
Moreover, we are interested in evaluating the impact of the attacks on the Ardupilot failsafe mechanism (see Section~\ref{sec:priordetectors}). For this reason, we measure how often the attack triggers the EKF failsafe, how often the drone crashes, how often the drone completes the mission, and how often it performs an emergency landing. For the attacks that trigger the EKF failsafe or crash the drone, we compute respectively the average Time to Detect (TtD) and Time to Crash (TtC). We note that with the Ardupilot default EKF failsafe the minimum Time to Detect is $1$ second (see Section~\ref{sec:priordetectors}).
\begin{equation}
    TtD=T_{\text{failsafe}}-T_{\text{attackStart}}
\end{equation}
\begin{equation}
    TtC=T_{\text{crash}}-T_{\text{attackStart}}
\end{equation}
\begin{table}[tb]
\caption[Behavior characterization of Ardupilot reaction to \attacks]{Emulation of \attacks with Ardupilot SITL. Behavior characterization of Ardupilot flight control after launching \attacks for $1$~second. Each attack is repeated $11$ times.}
    \label{tab:attack_impact}
    \centering
    \resizebox{\columnwidth}{!}{\begin{tabular}{rrrrrrrrr}
    \toprule
&\multicolumn{4}{c}{SDA behavior}\\
\cmidrule{2-5}
&\multicolumn{2}{c}{Undetected}&\multicolumn{2}{c}{Detected (failsafe)}&\multicolumn{2}{c}{TtD}&\multicolumn{2}{c}{TtC}\\
\cmidrule(l){2-3} \cmidrule(l){4-5}   \cmidrule(l){6-7}  \cmidrule(l){8-9}
{}
& \%crash & \%compl. & \%crash&\%land & $\mu$(s)&  $\sigma$ &  $\mu$(s) &  $\sigma$\\
\midrule
Stale
& 0.00   & 0.00  & 0.00 & 100.00 & 4.99 & 1.80 & N.A.  & N.A. \\
Default
& 0.00 & 0.00  & 100.00 & 0.00   & 1.22 & 0.00 & 4.56 & 0.00\\
Erron.
& 45.46  & 27.27 & 27.27 & 0.00   & 1.76 & 0.47 & 5.66 & 2.96 \\
Absent
& 100.00 & 0.00  & 0.00 & 0.00   & N.A.  & N.A.  & 4.22 & 0.01\\

\bottomrule
\end{tabular}}
\end{table}

\begin{table*}[t]
\centering
\caption[\attacks induced trajectory deviation.]{Emulation of \attacks with Ardupilot SITL. \attacks induced trajectory deviation. Attack duration is $1$~second. Each attack is repeated $11$ times.}
\label{tab:deviation}
\begin{tabular}{rrrrrrrrrrrrr}
    \toprule
        &\multicolumn{12}{c}{Maximum Deviation in meters until failsafe or crash}\\
        &       \multicolumn{4}{c}{x in meters} & \multicolumn{4}{c}{y in meters} & \multicolumn{4}{c}{z in meters}   \\
           \cmidrule(l){2-5}  \cmidrule(l){6-9} \cmidrule(l){10-13}
    &        \multicolumn{1}{c}{Max} &         \multicolumn{1}{c}{Min} &  \multicolumn{1}{c}{Mean} &         \multicolumn{1}{c}{Std} &  \multicolumn{1}{c}{Max} &         \multicolumn{1}{c}{Min} &\multicolumn{1}{c}{Mean} &         \multicolumn{1}{c}{Std} &  \multicolumn{1}{c}{Max} &         \multicolumn{1}{c}{Min} &  \multicolumn{1}{c}{Mean} &         \multicolumn{1}{c}{Std}   \\
\midrule
Stale
&34.46 & 0.54 & 16.67 & 15.82 & 17.24 & 0.03 & 7.60 & 7.86 & 2.62 & 0.09 & 1.24 & 1.17\\
Default   & 4.51  & 3.89  & 4.30  & 0.23  & 39.65  & 39.65  & 39.65  & 0.00  & 4.88  & 4.85  & 4.86  & 0.01 \\

Erroneous & 98.83 & 0.44 & 29.72 & 28.02 & 125.12 & 0.68 & 24.96 & 35.11 & 54.05 & 8.43 & 32.00 & 17.93\\
Absent    & 58.26 & 57.24 & 57.76 & 0.37  & 138.14 & 138.11 & 138.13 & 0.01  & 53.12 & 53.07 & 53.10 & 0.02 \\
\bottomrule
\end{tabular}
\end{table*}

\subsubsection{Emulation Results}
\label{sec:emulation_results}
Table~\ref{tab:attack_impact} and Table~\ref{tab:deviation} report the summary of the emulation carried out over the four \attacksShort. We measure the deviation induced by the \attackShort before the EKF failsafe triggers.

\Par{Stale Data} In the case of Stale Data attacks, the attack triggers the EKF failsafe $100\%$ of the times, which successfully lands the drone $100\%$ of the times. The EKF failsafe requires an average TtD of $5$ seconds which is a large amount of time (recalling that the minimum time to detect is $1$ second). The attack deviates the drone from its trajectory, $1$ to $16$ meters on average. The standard deviation shows that the attack-induced behavior depends on the value used for the stale data. Each emulation run is not deterministic as there is simulated noise in the sensor readings, consequently, the stale data will have a different value, causing different controller behavior.

\Par{Default Value Data}
In this case, the \attackShort-induced behavior triggers the EKF failsafe, but this time the drone crashes in every emulation as the controller cannot stabilize the vehicle after the attack. The induced behavior is consistent among each repeated experiment as we can observe a low standard deviation. This is explained by the fact that the transmitted data is always the same between runs. TtD is low, as the EKF failsafe is triggered consistently in $1.22$s (vs. $1$s minimum detection time).

\Par{Erroneous Data} In this case, the received data at the controller follows a normal distribution. This causes different behaviors in the drone as each sample received during the attack is random. The attack triggers the EKF failsafe in the $27\%$ of the cases (meaning that the attack is undetected most of the time). The times that the anomaly is detected, the drone crashes. When the attack is not detected, the drone crashes in the $45\%$ of the cases and recovers the $27\%$ of the times. When the anomaly is detected the detection requires on average $1.76$ seconds. The impact on the controller causes deviations that span from $24$ to $32$ meters on average.

\Par{Absent Data} In this case, the controller stops receiving sensor updates. In our emulations, the drone crashes $100\%$ of the time without triggering the EKF failsafe. The induced behavior is consistent among the runs (i.e.,~a low standard deviation of the induced maximum deviation). Notably, the attack will induce a deviation of $138$ meters on the y axis and above $50$ meters on the x and z axis.

In Appendix~\ref{apx:distribution} we show the distribution of the TtD for the different sensor behaviors in reaction to the \attacksShort.

\begin{formal}

    About \textbf{\hyperref[rq3]{RQ3}}, the emulation results of \attacksShort show the severe impact of such manipulation on the physical process, which causes drone deviation from the expected path and crash. Moreover, we characterized the behavior of the EKF failsafe mechanism in reaction to the \attacksShort. Our results show that Stale data attacks require a longer time to detect and the drone can safely land after detection. Default data attacks are quickly detected, but the drone will crash. Erroneous data are undetected most of the time and the induced behavior varies (most of the drone crashes). Absent data are always undetected and cause drone crashes. In Section~\ref{sec:synthesis}, we propose an attack synthesis framework to fly the drone to an adversarial position.
\end{formal}

\section{Physical Deviation: Attack Synthesis}
\label{sec:synthesis}
To solve the adversarial control problem identified in Equation~\ref{eq:adversarial_control}
we propose a framework for \attackShort controllability based on Reinforcement Learning (RL). The proposed framework allows the attacker agent to deviate a drone toward an attacker's desired location. To reach the desired location, the attacker decides when the \attackShort should be performed each time ($\gamma_k \in \{0,1\}$, see Section~\ref{sec:attack_formalization}).

The attack synthesis relies upon the following elements:

\Par{Goal} The goal is an X, Y, and Z coordinate to which we aim to deviate the drone. We randomize the goal at the beginning of each episode to guarantee the generalizability of the learned policy.

\Par{Action Space} The action space is a boolean variable, reporting whether the IMU readings should attacked.

\Par{State Observations} The observation space is 25 dimensional and continuous, with X, Y, and Z drone position in meters, quaternion orientation (4 values), Roll, pitch, and yaw angles in radians, the velocity vector in m/s (3 values), Angular velocity in radians/second(3 values), Motors' speeds in RPMs (4 values), the roll, pitch and yaw first derivative and X, Y, Z goal position of the goal.

\Par{Reward Function} We design a reward function to train the RL agent for our drone deviation task (see Algorithm~\ref{alg:reward} in Appendix~\ref{apdx:reward}). The reward function is designed to solve the dual of the adversarial control problem (see Equation~\ref{eq:adversarial_control}). The reward function evaluates the Euclidean distance to the goal (\verb|goalDist|) and the change of the drone distance to the goal ($\Delta$\verb|Dist|). Moreover, it rewards the drone flying properly (\emph{droneUP()}), and penalizes it if upside down or flipping.  Finally, it penalizes the drone for following the controller's desired path (i.e.,~not flying away from the planned trajectory), penalizes it if the Ardupilot EKF failsafe is triggered, and awards it when gets closer to the goal.

\subsection{Attack Synthesis Framework}
In the previous sections, we introduced the \attacksShort and evaluated their impact on drone operations and COTS flight controllers. In this section,  we investigate the following questions \label{rq4}\textbf{RQ4}:
\emph{How can UAVs be adversarially controlled via \attacksShort?} To answer this question, we investigate the controllability of the proposed \attacksShort and propose an attack synthesis methodology to fly the drone via \attacksShort.
\begin{figure}
    \centering
    \includegraphics[width=0.8\columnwidth]{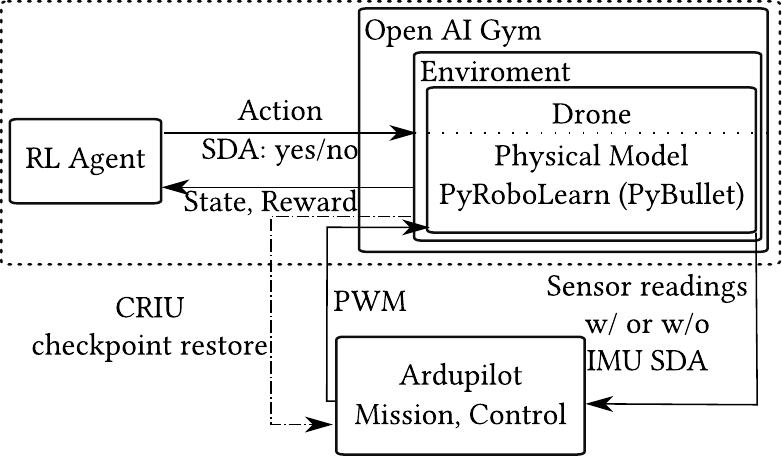}
    \caption{Controllable attack synthesis framework}
    \label{fig:rl-framework}
\end{figure}

\Par{Framework Implementation} Figure~\ref{fig:rl-framework}, summarizes the framework architecture for our evaluation. We create an Open AI Gym~\cite{gymenviroment} environment to allow the interaction between the Reinforcement Framework agent and Ardupilot. Ardupilot SITL allows for an external physical emulation~\cite{externalemulation} via UDP socket, we rely on this feature to enable the interaction between the reinforcement learning environment and Ardupilot. To initialize the Ardupilot controller in the environment, we use Checkpoint/Restore In Userspace (CRIU) to restore the Ardupilot process. The Ardupilot image was taken while the drone was flying a mission, and following a straight line.

Learning is based on episodes. Each episode is composed of steps. In each step, the agent computes the \attackShort decision based on the last received state and transmits it to the environment. The physical model will receive the PWM input from Ardupilot, simulate the physical response, send to Ardupilot the updated sensor reading or the attacked sensor readings (based on the attack decision from the agent), and send the updated physical state to the agent, which starts the next step of the framework.

\begin{figure}[tb]
  \centering
  \includegraphics[width=\columnwidth]{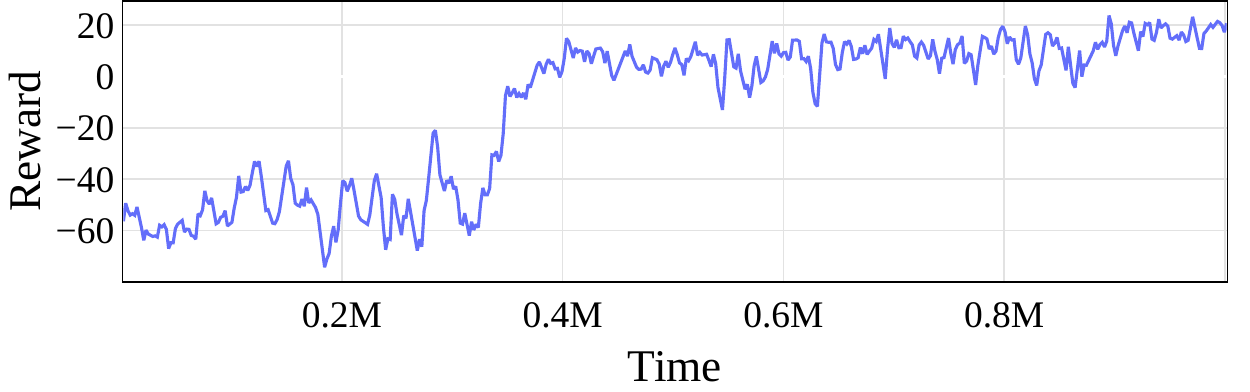}
  \caption{Attack synthesis: training reward, 1M iterations}
\label{fig:reward}
\end{figure}

\Par{Evaluation of the attack synthesis framework} We train the attack synthesis agent for 1M iterations (19 hours) using the TRPO algorithm~\cite{schulman2015trust}, Figure~\ref{fig:reward} reports the reward evolution over the evaluation set (tested every 10k iterations). As we can see the model over time increases the average reward function value, which means that the agent policy learned how to deviate the drone via \attacksShort.

\begin{figure}[tb]
  \centering
  \includegraphics[width=\columnwidth]{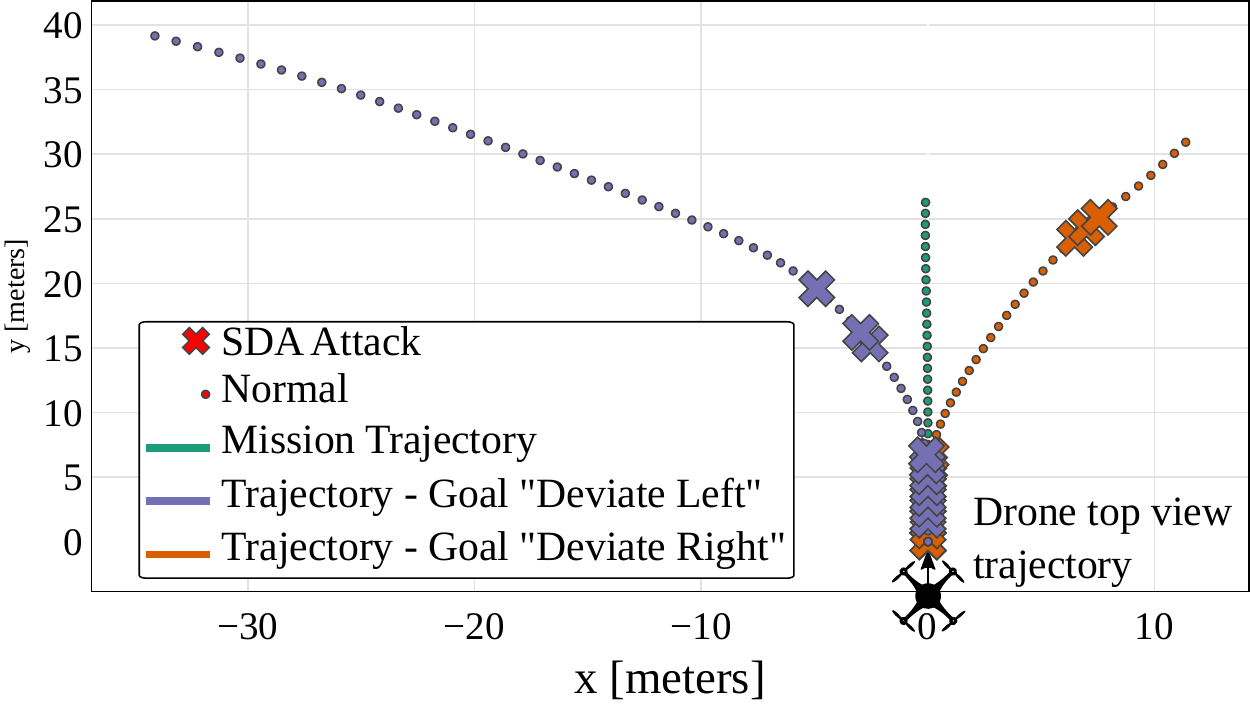}
  \caption{Drone Traces, the attack synthesis learned policy. Given the objective, the drone will turn left or right via intermittent SDAs}
\label{fig:DroneTraces}
\end{figure}

We evaluate the trained agent's capability to perform the learned task. Figure~\ref{fig:DroneTraces} reports sample traces for the learned policy for the deviating left/right task, compared to the Ardupilot planned trajectory. The drone is flying on a straight trajectory and would succeed in following the trajectory unless \attacksShort are performed.  The figures show the steps where the synthesis framework sent \attackShort action to achieve the desired goal. The results show that the proposed attack synthesis methodology can be used to strategically launch \attacksShort to adversarially fly the drone away from the planned trajectory, towards an attacker's desired location without triggering the failsafe.

\begin{formal}

To answer \textbf{\hyperref[rq4]{RQ4}} we propose an attack synthesis methodology based on reinforcement learning. Our evaluation results show that the target UAV can deviate from the planned trajectory to the attacker's desired location. Compared to prior work attacks, the proposed framework represents an advancement in the attacker capability as previously demonstrated attacks will imply drone crashing~\cite{son15rocking,jang23paralyzing}.

\end{formal}

\section{Defenses}
This section discusses defenses against \attacksShort, we show that state-of-the-art UAV anomaly detection techniques such as M2MON~\cite{khan2021m2mon}, fail to detect \attacksShort. Moreover, we discuss defenses from \attacksShort, at the software and system level.

\subsection{Detecting \attacksShort}
\label{sec:detection}
In previous sections, we present and evaluate a novel attack vector against UAVs. We are now interested in understanding if and how the proposed \attacksShort are detected by the state-of-the-art attack detectors for UAVs. In particular, we are interested in answering this research question \label{rq5}\textbf{RQ5}: \emph{Are the induced \attackShort detected by State-of-the-Art detectors for UAVs?}

In Section~\ref{sec:emulation}, we assessed the impact of \attacksShort on the EKF failsafe algorithm implemented on Ardupilot, showing that the attacks are often undetected. To answer \textbf{\hyperref[rq5]{RQ5}}, we are now interested in assessing the impact of \attacksShort on UAV anomaly detectors proposed in prior work. We evaluate \attacksShort on flight controllers with the state-of-the-art security reference monitor M2MON \cite{khan2021m2mon} (see Section~\ref{sec:priordetectors}). To detect an attack, M2MON continuously monitors the MMIO (Memory-mapped I/O) address range of the MCU with 3 different metrics:
\emph{Access list} (list of MMIO address access),
\emph{Access chain} (order of MMIO address access), \emph{Access frequency} (frequency of MMIO address access). Any deviation from regular access patterns indicates the presence of malicious actors. Additionally, to secure the EKF from attacks on the RTOS, M2MON runs EKF separately from the RTOS. Since full M2MON implementation is not available for the current ArduPilot project, we re-implement these monitors. Then, we show that \attacksShort can remain undetected or delay detection.

\Par{Monitoring MMIO Access}
The Kakute H7 Mini (with STM32H743 MCU and BMI270 IMU) uses the SPI interface for MCU-to-IMU communication. Upon referencing the datasheet of STM32H743 we find that the transmit data register (TXDR) for SPI is located at address \verb|0x40013420| and the receive data register (RXDR) for SPI is located at address \verb|0x40013430|. These registers are responsible for exchanging data with the IMU. Therefore, we inject counters into low-level SPI drivers for the ArduPilot firmware and count the number of accesses for the specified SPI MMIO addresses.

\Par{Anomaly Detection} \attackShort does not alter the list and order of MMIO address access. As the \attackShort does not modify the control firmware, the MCU will continue querying the IMU. Therefore, the \attackShort remains undetected under the first two M2MON metrics, i.e., access list and access chain. Further, we investigate the change in the frequency of MMIO address access. To achieve this, we first calculate the maximum and minimum MMIO access frequency for $300$ seconds during the regular operation of IMU. To validate if MMIO access always remains within the calculated boundary, we compare it with a different $60$ seconds trace of regular IMU operation. The detector has an accuracy of $0.999988$ when monitoring idle MCU.
We observe that the regular operation of IMU suffers from transient changes in access frequency and this leads to a little number of false positives even when there is no attack happening.

Finally, we perform sensor suspend
on our IMU for $60$ seconds and use the previously derived boundaries to detect \attacksShort.

\emph{Sensor Suspend Attack Detection.} In this case,
the detector has a low accuracy of $0.0002$ when applied on suspend attacks. For this reason, we argue that this attack remains undetected. We attribute that sensor suspend attacks remain undetected by M2MON metrics since there is no change in communication patterns with the BMI270 IMU. Moreover, monitoring the values that are set into the registers will not be sufficient to detect the device reconfiguration attack, the attacker issues the reconfiguration through an Out of Bound (e.g.,~IEMI) and not through the monitored MCU registers.

\emph{Process-based detection.} Similarly to what we showed in Section~\ref{sec:emulation} about the impact of SDAs on the EKF failsafe mechanism, we argue that the time to detect increases also in the case of other proposed process-based anomaly detectors (e.g.,~Control Invariant~\cite{choi18CI} and SAVIOR~\cite{quinonez20savior}) which rely on the same sensor observation used by the failsafe to detect anomalies.

\begin{formal}
  To answer \textbf{\hyperref[rq5]{RQ5}}, we evaluate the proposed attacks against prior work anomaly detectors and show that the proposed suspend attack evades M2Mon (accuracy drops  from $0.999988$ to $0.0002$) and induce delays in the process-based anomaly detection.
\end{formal}

\subsection{Defending Against \attacksShort}
The proposed attacks work by changing the configuration of the sensor chip. Additionally, as we explained in Section~\ref{sec:hardware}, sensor configuration and reconfigurations occur via an unauthenticated bus. To the best of our knowledge, the hardware designs of sensor chips do not provide secure configuration options, which rules out any authenticated bus option. Mitigating the proposed attacks at the sensor chip level would require: i) To enable the chips' secure configuration, there will be a need for an authentication layer on the serial protocol. This is challenging as it induces computational and power overheads on resource-constrained devices and it cannot be retrofitted to legacy chips. ii) To remove configuration functionalities from the sensors. This way, the device configuration cannot be overwritten by an attacker (losing the freedom of configuration that enables general-purpose deployment of sensors). Alternatively, allow configuration only at sensor boot. By doing so if an \attackShort occurs, either it occurs at boot and consequently impedes the drone arming, or it would not be possible to launch it during flight.

Software level mitigations would not take away the root issue of sensor reconfiguration but are easier to deploy in legacy hardware. Periodical attestation of the sensor configuration might help mitigate the attack. Another possibility would require the ability to monitor the bus for anomalous requests (both at the controller and at the sensor side). Those solutions will introduce traffic overhead which might contrast with the application's real-time requirements.

\Par{Attack Recovery} As we shown in our evaluation, state-of-the-art attack detection is evaded by our proposed strategy this implies that recovery would be not possible in general as the trigger for attack recovery is the attack detection. Assuming that the reconfiguration attack is detected, reissuing the correct configuration commands would be insufficient, as the attacker could reconfigure the victim. Khan et al.~\cite{khan2021m2mon} proposed to open the parachute as soon as an attack is detected. Choi et al.~\cite{choi20recovery} proposed using software sensing techniques to recover from attacks in RVs. Dash~\cite{dash2024diagnosis}, proposed an attack diagnosis-based UAV recovery strategy. We argue that the recovery strategy depends on the context where the RV is deployed. Parachute opening is a conservative solution, while software sensing might allow drone recovery. If software sensing fails, opening the parachute will be the fallback solution.

\section{Related Work}
\label{sec:related}
We discussed relevant related work in Section~\ref{sec:background}. In this section, we report complementary related works.

\Par{Control in Presence of Missing Deadlines} In control theory, the problem of control in the case of missing observations or missing control commands has been formalized and studied. Sinopoli et al.~\cite{sinopoli2004kalman} lay the foundations for this area of research, studying the convergence of Kalman Filter-based estimation in case of network faults. Their results show an upper limit of missing observations above which the state covariance matrix of the filter becomes unbounded and makes state estimation unreliable. Schenato et al.~\cite{schenato2007foundations, Schenato2009tozerotohold} studied the problem of controllability in the case of lossy network communication and the performance of zero-input and the hold-input scheme control strategies. More recently, Maggio et al.~\cite{maggio2020control} studied the stability guarantees of real-time control systems under sporadic deadline misses. Related works study faulty behaviors of the control network with a stochastic approach, in our case the missing observations are attacker-intentional.
Finally, Li et al.~\cite{li2024empirical} present TimeTrap, software fault injection technique to find exploitable timing patterns in control software, and determine optimal strategy to maximize the control deviation.

\Par{Adversarial Reinforcement Learning}
Recently, Reinforcement Learning techniques were applied to security-related contexts in attack and defenses. On the defensive side, Landen et al.~\cite{Landen22Dragon} recently proposed a deep reinforcement learning agent to operate the power grid and perform anomaly detection, while Maiti et al.~\cite{maiti2023targeted}, proposed a reinforcement learning agent to perform vulnerability assessment of the power grid.  On the attack side, Wu et al.~\cite{Wu21adversarialpolicy} and Gong et al.~\cite{Gong22curiosity} proposed policies to train deep reinforcement learning adversarial agents to attack state-of-the-art reinforcement learning agents. We are the first to investigate the use of reinforcement learning to adversarially perform control of a victim.

\section{Conclusion}
\label{sec:conclusion}

In this work, we presented novel \attacks,  exploiting sensor reconfiguration to achieve process destabilization. The proposed attacks differ from prior work attacks on UAVs (e.g.,  GPS spoofing) as the attacker does not need to continuously spoof sensor readings to achieve the goal. The proposed attack is practically demonstrated and evaluated on COTS flight controllers. We extensively investigate and describe the effects of the attack at the hardware and control level, showing that the attacks severely delay the drone attitude estimation since EKF estimation frequency depends on IMU frequency. Moreover, we demonstrate how to use \attacksShort to control the victim's vehicle. We show that the proposed attack can evade state-of-the-art UAV anomaly detection techniques (accuracy drops from $0.999988$ to $0.0002$) or delay attack detection while deviating the victim vehicle more than $30$ meters from the planned trajectory.
Our findings demonstrate how the proposed \attacksShort threaten modern UAVs and their consequences on state-of-the-art control software. To defend against such attacks, we discussed possible countermeasures that range from modification to chip design to software-level mitigation.

\bibliographystyle{acm}
\bibliography{bibliography}
\appendix

\section{Alternative SDA attack: Frequency Attack}
\label{sec:frequencyAttack}

To launch an SDA, an alternative to sensor suspension is reconfiguring the sensor to operate at a lower sampling frequency. Since the sensors commonly found on modern flight controllers are general-purpose devices (e.g.,~the same IMU chip can be used in various contexts, for example, in smartphones, activity trackers, etc.) they can be programmed for different purposes. The sensor's sampling frequency is one feature that can be reconfigured to serve a particular task. In flight control software, sensors are often configured to run at the highest sampling frequency, this allows precise control of the aircraft. An attacker can resort to this device configuration to decrease the update rate of the sensor, in this way the sensor readings are updated less frequently, and the controller will rely on outdated sensor readings or wait until the next sample is collected. The frequency reconfiguration can be modeled as an intermittent suspend reconfiguration where the suspension lasts $\frac{1}{\text{sensor frequency}}$.

\begin{figure}
    \centering
         \includegraphics[width=\columnwidth]{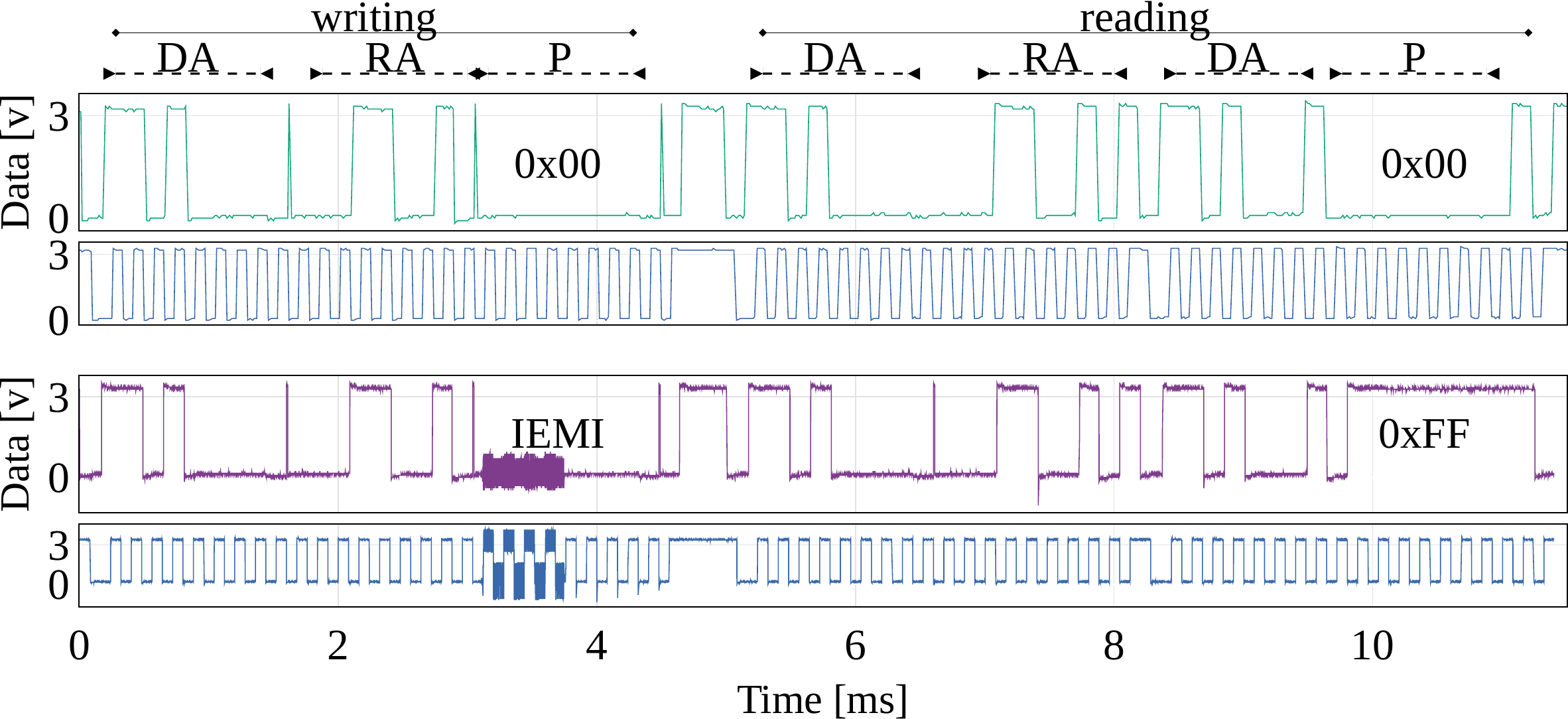}
         \caption{Sensor sampling frequency modification via IEMI attack on MPU6050, the IEMI modifies the payload of the message from 0x00 to 0xFF.\label{fig:MPU6050frequency}}
\end{figure}

\subsection{Reconfiguration via IEMI}
To achieve the sensor sampling frequency change in the MPU6050, we have to write in the \verb|sampling rate divider| register \verb|0x19|. The value contained in this register will be later used to determine the sampling rate as in Equation~\ref{eq:sample}.
\begin{equation}
    \text{Sample Rate} = \frac{\text{Gyroscope Output Rate}}{ (1 + \text{Sampling Rate Divider})}
    \label{eq:sample}
\end{equation}
Figure~\ref{fig:MPU6050frequency}, represents the collected power trace at the oscilloscope, with and without injection. As we can observe when the IEMI injection is performed (lasting 640$\mu$s, i.e., 4 clock cycles), the response from the register changes, and the value that is stored in the register is \verb|0xFF|, which is the highest 8-bit unsigned integer, corresponding to the lowest sampling frequency configurable ($31.25$Hz).

\begin{figure*}
  \centering
\subcaptionbox{BMI270\label{fig:BMI270MainLoop}}{\includegraphics[width=0.49\linewidth]{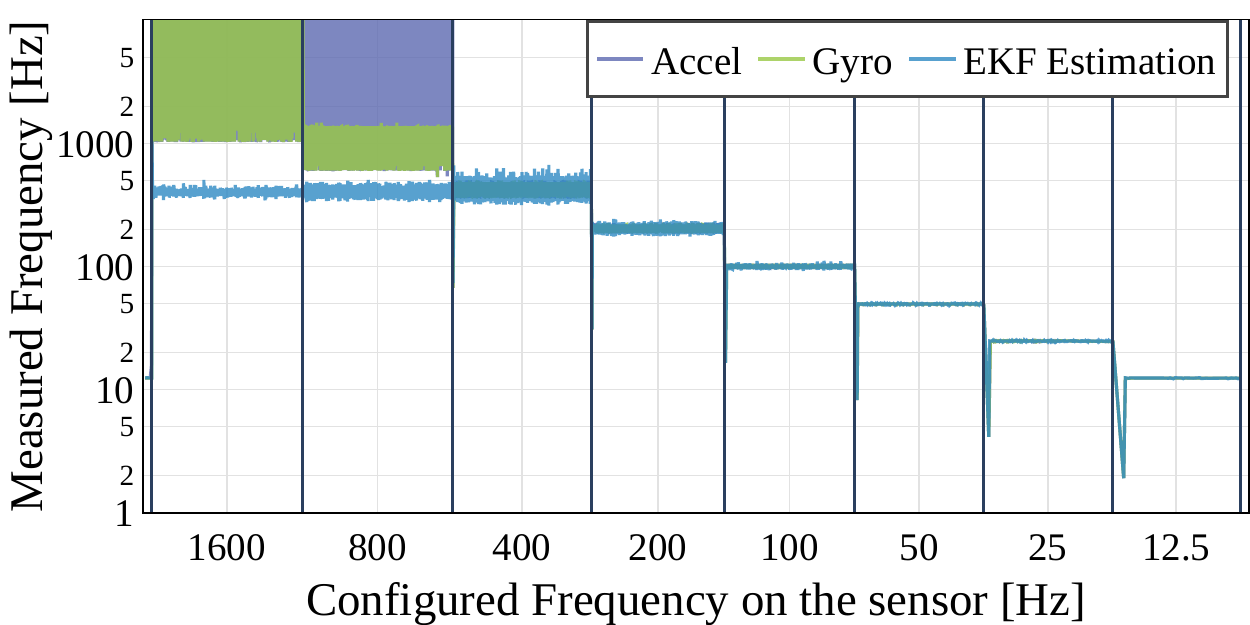}}
  \subcaptionbox{BMI055\label{fig:MainLoopBMI055}}
  {\includegraphics[width=0.49\linewidth]{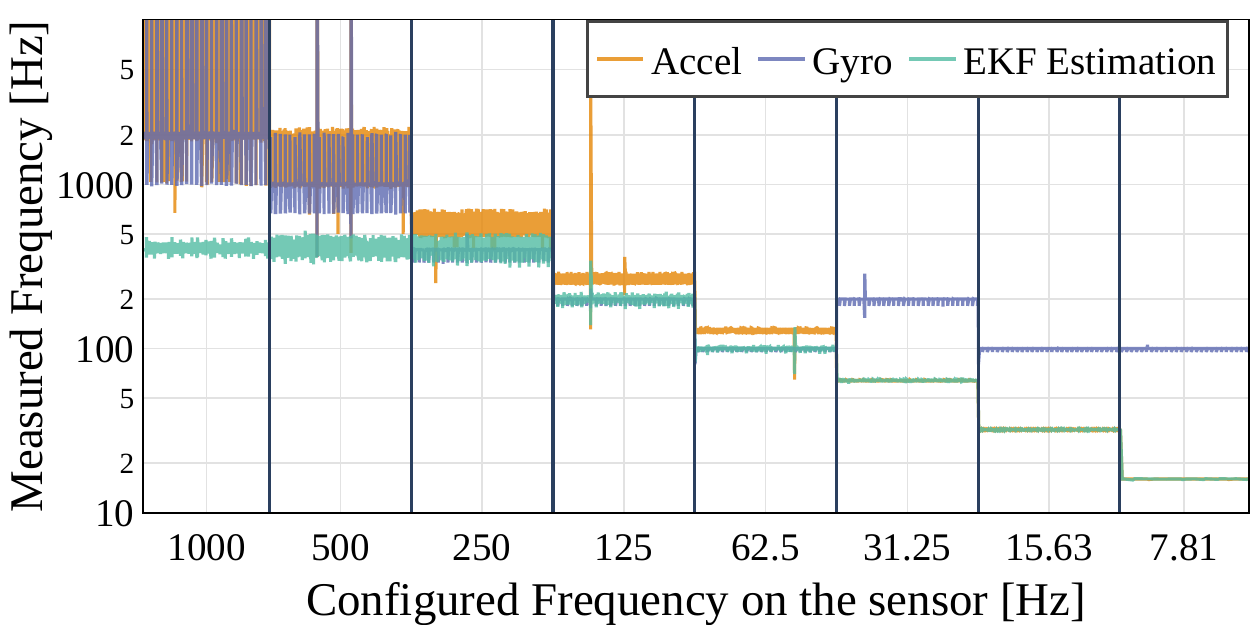}}
\subcaptionbox{ICM42688\label{fig:MainLoopICM42688}}
  {\includegraphics[width=0.49\linewidth]{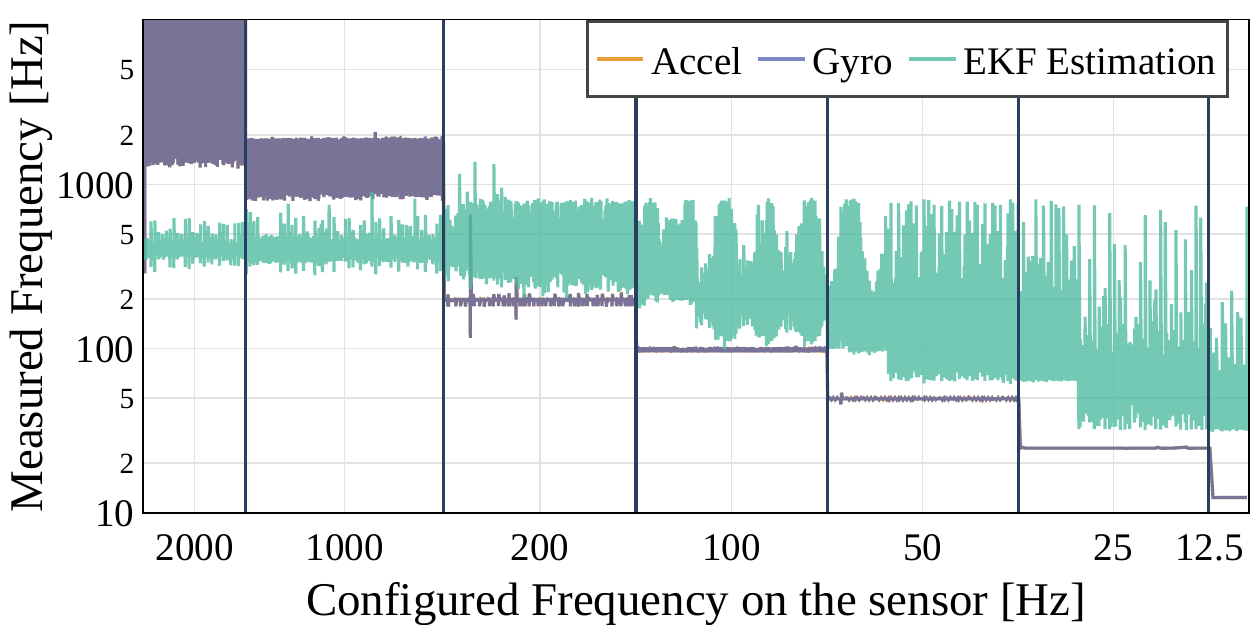}}
\subcaptionbox{MPU6000\label{MainLoopMPU6000}}
  {\includegraphics[width=0.49\linewidth]{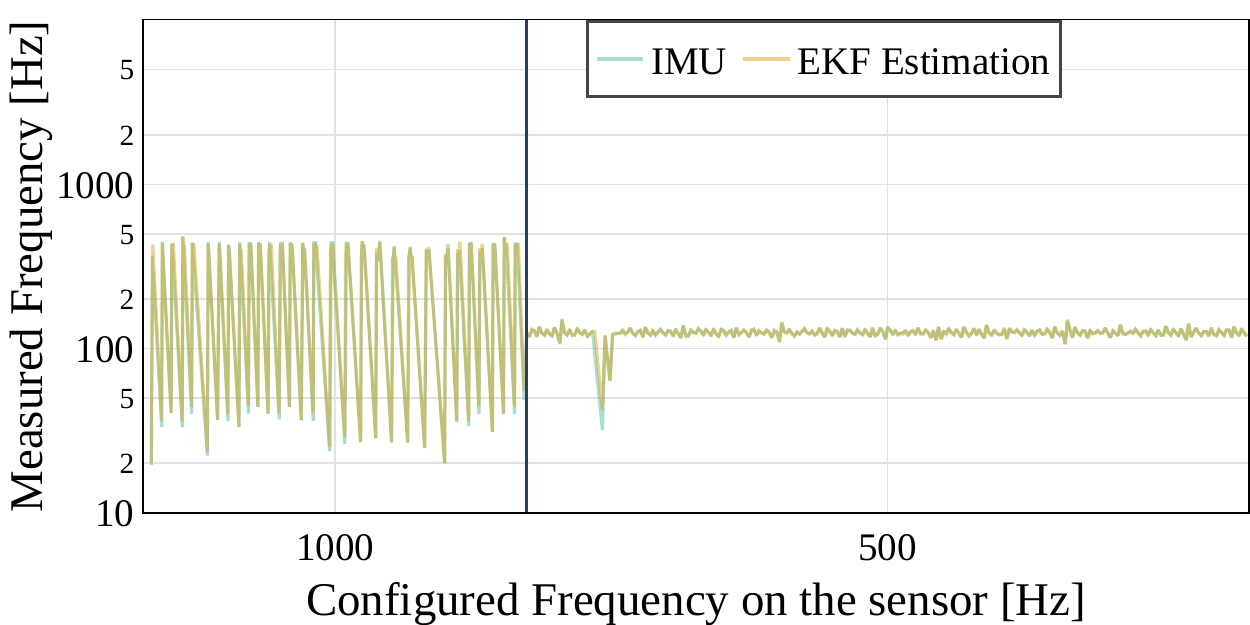}}
\caption{ Dependency between the IMU sampling frequency and the attitude estimation frequency. Below 400Hz, sensor frequency slows down to the state estimation.}
\label{fig:MainLoopEffects}
\end{figure*}

\begin{table}[h]
    \centering
        \caption{Summary of IMU behaviors observed during IMU frequency attack}
    \begin{tabular}{rrlrl}
        \toprule
        &   \multicolumn{4}{c}{ Lowest frequency (Hz)} \\
        &   \multicolumn{2}{c}{Accelerometer } & \multicolumn{2}{c}{Gyroscope}\\
        IMU Model    & Tested  & Allowed & Tested & Allowed\\
        \midrule
        BMI055 & 15.56 & 7.81 & 32 & 32\\
        BMI270  & 12.5 & 0.78125 & 25 & 25\\
        ICM-42688-P  & 12.5 & 1.5625 & 12.5 & 12.5\\
        MPU6000  & 100 & 50 & 100 & 50\\
        \bottomrule
    \end{tabular}

    \label{tab:frequencies}
\end{table}

\subsection{Effects of IMU Sampling Frequency}
To test the effect of frequency change on the control software (Ardupilot), we set all the possible allowed sampling frequencies in the target IMU. The sensor is by default configured to operate at the highest sampling frequency (e.g., for the BMI 270, 3200Hz for the gyroscope, and 1600Hz for the accelerometer). Then, we reconfigure the sensor to operate at a lower frequency. Figure~\ref{fig:BMI270MainLoop}, shows the impact of the frequency change compared to the Kalman Filter estimation frequency of the system when performing experiments on the the four considered IMUs that show consistent behavior. In Ardupilot, the Kalman Filter prediction is scheduled for computation at 400Hz. As long as the sensor sampling frequency is higher than 400Hz, the attack does not show any consequence on the controller. When the sensor frequency is reduced below 400Hz, we can observe that the Kalman Filter estimation gets slowed down. This has severe consequences on the scheduler and drone control. In a cascading effect, the drone attitude estimation will not produce fresh outputs and consequently, no control commands are computed.
Moreover, we observed scheduled tasks in the ROTS being slowed down due to deadline misses (e.g., data transfer to the ground station will slow down and barometer sensors will be sampled less frequently). In the next paragraph, we comment further on the observed consequences of our \attacksShort.

In our experiments, for all the considered IMUs, we were able to slow down the sampling frequency impacting the Kalman Filtering operations. Table~\ref{tab:frequencies}, reports the lowest possible frequencies that can be set on each IMU.

\Par{Explanation for observed delay effects} Given the severe consequences of IMU sampling frequency on EKF frequency, and other sensors (e.g., the barometer) we investigated the root cause of this behavior by looking into the ArduPilot codebase. We observed that the scheduler loop is running continuously inside an infinite while loop. The scheduler loop waits for new data from IMU and manages 81 tasks (such as the EKF state estimator, attitude controllers, and checking if the UAV has landed or crashed). Differently from what is expected, although the reading of data from IMU is performed asynchronously, when the \verb|sensor frequency |$<$\verb| scheduler frequency| the scheduler loop waits for new IMU data to continue its execution. As a consequence, all the tasks are not performed until a new reading is obtained from IMU.

Our \attackShort exposes a controller design flaw, which makes it possible to delay the tasks inside the firmware from a ROTS execution inside the MCU from an untrusted sensor.

\Par{Lowest frequency for control} Using the Ardupilot SITL simulator, we test the frequency change attack to verify the impact of the sensor frequency change on the flight and observe which is the frequency below which the flight controller will fail in controlling the drone. In our evaluation, we found that below 200Hz the drone becomes unstable, but continues the mission. Instead, below 150Hz the controller will fail to stabilize the drone and will crash on the ground. All the tested IMU allow frequencies lower than 150Hz.

\begin{formal}
    Interestingly, we discovered that the sensor sampling frequency attack will force the attitude estimation to slow down to the frequency of the sensor and impact the whole RTOS scheduler. This complements our previous findings for  \textbf{\hyperref[rq3]{RQ3}} and presents an effective manipulating alternative.
\end{formal}

\subsection{Sampling Frequency Attack Detection}

In this case, we observe that the detector has an accuracy of $0.9377$ for 100 Hz and $0.9923$ at 12.5 Hz re-configuration. This is because a sampling frequency attack leads to a decrease in communication frequency between the IMU and the MCU. We argue that while the detector would be able to observe a change in access frequency, this detection will be delayed by the lower frequency of the IMU as there are 128 missing consequent observations when changing the IMU frequency from 1600Hz to 12.5Hz. We note that in the case of Absent Data, all the observations will be missing. Moreover, M2MON-EKF would wait for subsequent IMU readings for state estimation delaying the overall operations of M2MON (32 skipped state estimations, i.e.,~12.5Hz vs 400Hz). As already noted by Jang et al.~\cite{jang23paralyzing}, induced higher time to detect the anomaly will consequently reduce the probability of attack remediation (failsafe/parachute) that leads the drone to crash. We note that since the implementation of M2MON-EKF is the same EKF implementation in ArduPilot, moving EKF implementation outside RTOS offers no additional security against \attacksShort.

\section{Detection distribution of proposed attacks}
\label{apx:distribution}
\begin{figure}
    \centering
    \includegraphics[width=0.7\columnwidth]{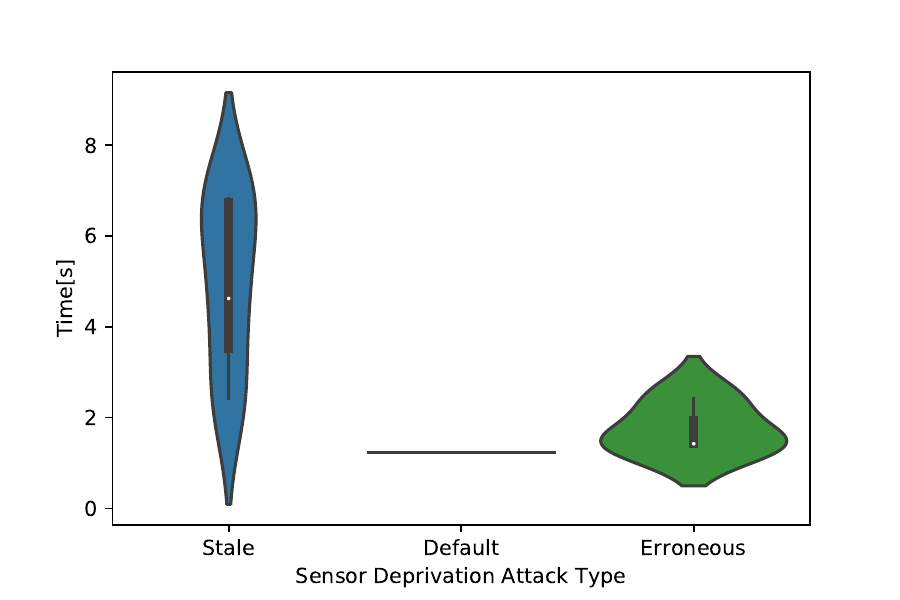}
    \caption{Time to Detect: violin plot representing the distribution.}
    \label{fig:violin_ttd}
\end{figure}

We evaluate when the EKF failsafe mechanism was triggered by the effects of any of the \attacksShort. Figure~\ref{fig:violin_ttd} represents the distribution of the time-to-detect of the failsafe mechanism. The Stale attack shows a wider distribution, reaching over 6 seconds of detection time after launching the attack.

\section{Reward function pseudocode}
\label{apdx:reward}

In this section, we present the reward function to train the RL agent to control the drone using one of the \attacksShort. In a nutshell, the reward function aims to generate a policy that `drives' the drone to a desired position (GoalPos) without losing control (reduced amount of flips) and without triggering any safety mechanism (failsafe). To generate a useful policy, the Reward function penalizes the trigger of the failsafe mechanism (line 22 in Algorithm~\ref{alg:reward}) and proportionally reduces the reward when the drone flips (line 16 in Algorithm~\ref{alg:reward}).
\onecolumn
\begin{minipage}[t]{0.45\columnwidth}
\begin{algorithm}[H]
\label{alg:reward}
\caption{Reward function}

\var{flips}  = \var{0}\\
\Function{Reward(state)}{
\var{reward} = \var{0}\\
\var{goalDist} = $\sqrt{(\var{DronePos} - \var{GoalPos})^2}$\\
\var{$\Delta$Dist} = \var{prevDist} - \var{goalDist}

\uIf{droneUP() and closeToGoal(\var{$\Delta$Dist})}{
        \var{reward} = \var{reward}+\var{1.5}
    }
    \uElse{
        \var{reward} = \var{reward}-\var{0.5}
    }

\uIf{!droneUP() and closeToGoal(\var{$\Delta$Dist})}{
        \var{reward} = \var{reward}+\var{0.5}
    }
    \uElse{
        \var{reward} = \var{reward}-\var{1.5}
    }

\uIf{droneFilps()}{
    \var{flips}  = \var{flips} + \var{1}

    \var{reward} = -\var{flips}
}
\uElse{
    \var{flips} = \var{0}
}
\uIf{closeToPlannedPath()}{
            \var{reward} = \var{reward} - \var{2}
}
\uIf{failsafeTriggered()}{
           \var{reward} = \var{-40}
   }
\uIf{\var{goalDist} < \var{1}}{
            \var{reward} = \var{100}
    }
    \Return{\var{reward}}
}
\end{algorithm}
\end{minipage}
\twocolumn

\end{document}